\documentclass[reprint,amsmath,amssymb,aps]{revtex4-2}
\usepackage[margin=0.75in]{geometry}
\usepackage[utf8]{inputenc}
\usepackage[english]{babel}
\usepackage{todonotes}
\usepackage{upgreek}

\usepackage{amsmath,amssymb,soul,subcaption,graphicx,color,lipsum,natbib,xcolor,bm,amsfonts,amstext,mathtools,amsthm,comment,ragged2e,helvet}
\usepackage[normalem]{ulem}
\usepackage[colorlinks=true, allcolors=blue]{hyperref}
\usepackage[font={normalsize}, skip=0pt,format=plain,singlelinecheck=false]{caption}
\begin{document}

\title{Epigenetic feedback reshapes dynamical landscapes in gene regulatory networks}

\author{Sascha H. Hauck$^{1,2}$}
\email{sascha.hannes.hauck@itwm.fraunhofer.de}

\author{Sandip Saha$^{3,4}$}
\email{sahasandip.loknath@gmail.com}
\thanks{Contributed Equally}
\thanks{Corresponding Author}

\author{Narsis A. Kiani$^{6,7}$} 
\email{narsis.kiani@ki.se}

\author{Jesper N. Tegn\'er$^{5,6,7,8,9}$} 
\email{jesper.tegner@kaust.edu.sa}

\address{$^1$Department of Flow and Material Simulation, Fraunhofer ITWM, Kaiserslautern, 67663 Germany}
\address{$^2$Chair for Scientific Computing, University of Kaiserslautern-Landau (RPTU), 67663 Germany}
\address{$^3$National Centre for Biological Sciences-Tata Institute of Fundamental Research, Bengaluru 560065, India}
\address{$^4$Biological and Environmental Science and Engineering Division, King Abdullah University of Science and Technology, Thuwal, 23955, Saudi Arabia}
\address{$^5$Division of Biomedical Sciences, King Abdullah University of Science and Technology, Thuwal, 23955, Saudi Arabia}
\address{$^6$Unit of Computational Medicine, Center for Molecular Medicine, Karolinska Institutet, Karolinska University Hospital, Stockholm, 171 76, Sweden}
\address{$^7$Algorithmic Dynamics Lab, Department of Oncology and Pathology, Karolinska Institute, Stockholm, 171 64, Sweden}
\address{$^8$Computer, Electrical and Mathematical Sciences and Engineering Division, King Abdullah University of Science and Technology, Thuwal, 23955, Saudi Arabia}
\address{$^9$Science for Life Laboratory, Solna, 171 65, Sweden}

\begin{abstract}
Understanding how gene regulatory networks (GRNs) give rise to stable and dynamic cellular states remains a central challenge in theoretical biology, particularly when slow epigenetic feedback reshapes the underlying regulatory landscape. While experimental approaches such as single-cell transcriptomics reveal rich dynamical behaviour, a tractable theoretical framework that links gene expression, epigenetic control, and collective dynamics remains challenging. Here, we develop an extended Dynamical Mean Field Theory (DMFT) framework for GRNs that incorporates epigenetic modifications as slow, feedback-driven variables. Building on the analogy between Hopfield networks and spin glass systems, we derive effective stochastic equations that reduce high-dimensional dynamics to a tractable form across multiple timescales. This formulation enables quantitative characterization of both stable and oscillatory regimes and reveals how epigenetic feedback reshapes the effective potential landscape governing cell fate decisions. Our model shows how epigenetic feedback regulation dynamically reshapes the Waddington landscape. Our results and methodology provide a unified theoretical framework for understanding developmental dynamics and epigenetic reprogramming in complex biological systems.
\end{abstract}

\maketitle

\section{Introduction}
\label{intro}
Understanding how high-dimensional biological systems generate stable yet flexible dynamical behavior is a central challenge at the interface of physics and biology~\cite{Waddington,tripathi2023minimal,Raju1,Raju3}. Gene regulatory networks (GRNs) govern cellular identity by orchestrating interactions among transcription factors, chromatin states, and signaling pathways~\cite{slack1991,Waddington,kaneko2007evolution,KanekoScience2012}. Yet, their dynamics are shaped not only by fast transcriptional processes but also by slower, history-dependent epigenetic modifications~\cite{kaneko2007evolution,KanekoPRR2020}. While experimental advances such as single-cell and spatial transcriptomics have revealed rich dynamical heterogeneity across developmental and disease contexts~\cite{Klein1,garg2023single,Mohit,Mohit2,wang2024transiently,Briscoe1,Briscoe2}, a general theoretical framework that links gene expression dynamics to epigenetic regulation across timescales remains lacking~\cite{KanekoPRR2020,matsushita2022dynamical}.

Classical approaches to GRN dynamics have drawn heavily on analogies to statistical physics, particularly through the connection between recurrent neural networks and spin-glass systems~\cite{raza2016recurrent,tripathi2023minimal,wang2024transiently}. In these models, stable gene expression patterns correspond to attractors of an energy-like landscape, providing an appealing interpretation of cell types as minima of an effective potential~\cite{Waddington,wang2024transiently}. However, such formulations typically assume symmetric interactions and equilibrium dynamics, thereby restricting the system to gradient flows that  converge to fixed points~\cite{Raju1,Raju2,Raju3}. This limitation is at odds with a growing body of evidence that biological systems operate far from equilibrium, exhibiting oscillations, lineage transitions, and metastable states that cannot be captured within purely dissipative frameworks\cite{slack1991,garg2023single,MuTrans,matsushita2022dynamical,Klein1,Briscoe1,Briscoe2}.

At the same time, epigenetic regulation introduces an additional layer of complexity by modulating the effective regulatory interactions over longer timescales~\cite{KanekoScience2012,kaneko2007evolution,KanekoPRR2020}. Chromatin accessibility, histone modifications, and DNA methylation dynamically reshape the regulatory landscape, enabling memory, hysteresis, and context-dependent responses~\cite{kaneko2007evolution,KanekoPRR2020,matsushita2022dynamical,matsushita2023generic}. These processes introduce feedback between gene expression and the underlying regulatory architecture, effectively coupling fast transcriptional dynamics to slow structural changes~\cite{kaneko2007evolution,miyamoto2015pluripotency,KanekoPRR2020,matsushita2022dynamical}. Despite their central role in development and reprogramming, such feedback mechanisms remain difficult to incorporate into existing theoretical models, which often treat regulatory interactions as static or assume a separation between dynamics and structure.

To address this gap, we develop a Dynamical Mean Field Theory (DMFT)~\cite{Sompolinsky1988,sompolinsky1982relaxational,crisanti1987dynamics} that encompasses GRNs with epigenetic feedback. DMFT provides a powerful approach for analyzing high-dimensional interacting systems by reducing their collective behavior to an effective stochastic process that captures both fluctuations and self-consistent interactions~\cite{Sompolinsky1988,sompolinsky1982relaxational,rajan2010stimulus,crisanti1987dynamics,van1992stochastic}. Originally developed in the context of disordered systems and spin glasses~\cite{Sompolinsky1988,sompolinsky1982relaxational}, and later applied to neural network dynamics~\cite{panda2017chaos,rajan2010stimulus}, DMFT is particularly well suited to studying large, heterogeneous networks with random or partially structured connectivity. Here, we generalize DMFT to incorporate slow, feedback-driven modifications of the interaction matrix, thereby coupling fast gene-expression dynamics to evolving epigenetic states.

Within this formulation, each gene is described by an effective stochastic differential equation whose statistics are determined self-consistently by the network~\cite{Sompolinsky1988,sompolinsky1982relaxational,rajan2010stimulus,panda2017chaos}. Epigenetic variables evolve on a slower timescale and modulate the strength and structure of regulatory interactions~\cite{KanekoPRR2020,matsushita2022dynamical}, thereby reshaping the effective potential landscape. This leads to a coupled system in which the landscape itself is no longer fixed but adapts to the system’s trajectory~\cite{KanekoPRR2020,matsushita2022dynamical}. As a result, the dynamics cannot be described solely in terms of convergence to static attractors; instead, they exhibit a rich interplay among stability, fluctuations, and the slow drift of the underlying regulatory structure.

Our analysis reveals how epigenetic feedback qualitatively alters the dynamical regimes of GRNs. In particular, we identify conditions under which stable attractors persist and regimes in which feedback induces oscillatory or metastable behavior. The framework allows us to characterize how memory and hysteresis emerge from the coupling between fast and slow variables, and how transitions between cellular states can be driven by gradual reshaping of the regulatory landscape rather than by external perturbations alone. Importantly, the DMFT reduction provides tractable equations for these phenomena, enabling analytical insight into systems that would otherwise be intractable due to their high dimensionality.

More broadly, this work establishes a unified theoretical perspective on gene regulation as a non-equilibrium dynamical system with evolving interactions. By integrating epigenetic feedback into a mean-field description, we bridge the gap between energy landscape models and time-dependent regulatory architectures, providing a framework that connects statistical physics, dynamical systems theory, and modern high-throughput biological data. This approach opens new avenues for understanding developmental processes, cellular plasticity, and epigenetic reprogramming in complex biological systems.

The study is organized as follows. In Sec.~\ref{model}, we introduce a mechanistic model for GRNs that incorporates an epigenetic term regulating and reprogramming the GRN dynamics through epigenetic feedback. We discuss the analytical challenges posed by this formulation and motivate a transformation to a new coordinate plane via a linear transformation, which facilitates analytical treatment and prepares the model for the MFT framework. This section also identifies the characteristic timescales governing the system dynamics. In Sec.~\ref{averagefield}, we apply the DMFT framework to the GRN model. This section focuses on the mathematical identification and approximation of the average field on longer timescales, as well as an approach for solving the slow-timescale dynamics required to compute autocorrelation functions and match dynamical moments. Sec.~\ref{autocorelation} is devoted to deriving the autocorrelation function under the assumption of clearly separable timescales. We highlight how this behaviour differs from the classical results of Sompolinsky~\cite{Sompolinsky1988} in the absence of epigenetic feedback or reprogramming terms. In Sec.~\ref{StabilityAnalysis}, we analyse the stability properties of the system and identify distinct stability regimes across multiple timescales. These regimes are characterized using the autocorrelation function within a Newtonian dynamics framework, obtained by mapping the microscopic network dynamics onto an effective macroscopic description. Finally, Sec.~\ref{Summary} summarizes the main findings of the study. We discuss the limitations of the proposed approach and framework and outline possible directions for future research.

\section{Mechanistic model of GRN–epigenetic interactions}\label{model}
During early development in multicellular organisms, cells differentiate into distinct types in a tightly regulated manner, ensuring stable cell fates despite noise or environmental fluctuations~\cite{matsushita2022dynamical,kaneko2007evolution,miyamoto2015pluripotency,KanekoPRR2020,Briscoe1,Raju3}. Waddington illustrated this robustness with the “epigenetic landscape,” where valleys represent stable cellular states. This landscape can be interpreted as a gene-expression dynamical system, in which cells move toward attractors within a regulatory network~\cite{matsushita2022dynamical,kaneko2007evolution,miyamoto2015pluripotency,KanekoPRR2020,Waddington,Raju3,Raju2}. Recent work has uncovered the molecular basis of epigenetic modifications, which contribute to the slow reshaping of this landscape~\cite{matsushita2022dynamical,kaneko2007evolution,miyamoto2015pluripotency,KanekoPRR2020}. Yet, how interactions between gene expression and epigenetic feedback drive this gradual remodelling-homeorhesis-remains poorly understood~\cite{matsushita2022dynamical,kaneko2007evolution,miyamoto2015pluripotency,KanekoPRR2020,KanekoScience2012}. Previous models suggest that epigenetic modifications adjust expression thresholds and, via positive feedback, reinforce cellular states, producing multiple attractors and hierarchical branching of valleys~\cite{matsushita2022dynamical,kaneko2007evolution,miyamoto2015pluripotency,KanekoPRR2020}. Our aim is to develop a quantitative framework based on DMFT to explain how such slow feedback processes drive homeorhesis, as originally envisioned by Waddington.

We consider a cell model that includes a GRN and epigenetic modification~\cite{matsushita2022dynamical,kaneko2007evolution,miyamoto2015pluripotency,KanekoPRR2020}. The cell contains $N$ genes, and its state is described by the expression levels $x_i$ of each gene $i$. Gene interactions are encoded in the regulatory matrix $J_{ij}$, where positive/negative values represent activation/repression. Gene expression follows an on/off-type response that saturates at high input levels. Normalizing the maximal expression to unity, we use the following dynamics:

\begin{align}
    \label{eq:ge_x}
    \dot{x}_i&= F\left(\sum_{j=1}^N J_{ij} x_j + \theta_i + c_i\right) - x_i\\
    \label{eq:ge_theta}
    \dot{\theta}_i &= \nu\left(\alpha x_i - \theta_i\right),
\end{align}
where $F(z)=\tanh(\beta z)$ with large $\beta(=40~\text{for instance})$, so that $x_i=1$ and $x_i=-1$ indicate full and no expression, respectively. The term $c_i$ represents an external input, typically set to zero. The parameter $-\theta_i$ acts as the effective threshold for gene $i$: increasing $\theta_i$ promotes expression, while decreasing it suppresses expression. Unlike standard GRN models with fixed thresholds, we treat $\theta_i$ as a dynamic epigenetic variable reflecting chromatin accessibility or histone modifications.

Our aim is to construct a DMFT framework to analyse the homeorhesis dynamics governed by Eqs.~\eqref{eq:ge_x} and \eqref{eq:ge_theta}. Classical DMFT was originally developed for spin glass systems~\cite{Sompolinsky1988,sompolinsky1982relaxational,crisanti1987dynamics,rajan2010stimulus,panda2017chaos} and later applied extensively to GRN models through Hopfield-type dynamics~\cite{Sompolinsky1988,crisanti1987dynamics,rajan2010stimulus,panda2017chaos}. However, the Hopfield framework does not typically account for slow regulatory feedback or epigenetic reinforcement~\cite{Mehta1,Mehta2}, making a direct application of standard DMFT inadequate for our purposes.

We turn to DMFT because it provides a powerful method--introduced by Sompolinsky--for reducing high-dimensional network dynamics to an effective one-dimensional description governed by Newtonian-like equations, thereby enabling tractable analysis of complex systems through local dynamics~\cite{Sompolinsky1988,crisanti1987dynamics,rajan2010stimulus,panda2017chaos,sompolinsky1982relaxational}. In our model, however, the presence of the epigenetic variable $\theta_i$ introduces an intrinsic feedback loop and slow timescale, complicating the direct use of conventional DMFT~\cite{Sompolinsky1988,crisanti1987dynamics,rajan2010stimulus,panda2017chaos,sompolinsky1982relaxational}.

Thus, we extend and adapt the DMFT methodology to handle this feedback-driven structure. By developing a mean-field formulation that separates the dynamics of the interacting components and leverages timescale separation, we aim to partially decouple the two coupled ODEs and obtain a tractable theoretical understanding of the full homeorhesis process~\cite{matsushita2022dynamical,kaneko2007evolution,miyamoto2015pluripotency,KanekoPRR2020}.

To construct such a MFT, we must carefully handle the activation function, as it inherently involves interactions between the network nodes. In the original MFT framework, these interactions are typically replaced by an effective noise term to reduce the global dynamics to a tractable form. However, for the homeorhesis model considered here, this replacement is non-trivial due to the feedback structure, making it necessary to decouple the activation function from the inter-node interactions. Accordingly, for the treatment of $x_i$, we introduce a new variable $y_i$ defined as,
\begin{align*}
    y_i = \sum_{j=1}^N J_{ij} x_j + \theta_i + c_i.
\end{align*}

Under the assumption that $c_i$ is constant, we can easily derive a ODE for $y_i$, similar to the above equation is given by
\begin{align}
\label{eq:ge_y}
    \dot{y}_i
    &= \sum_{j=1}^N J_{ij} \dot{x}_j + \dot{\theta}_i \nonumber \\
    &= \sum_{j=1}^N J_{ij} F\left(y_j\right) - y_i + \theta_i + \dot{\theta}_i + c_i.
\end{align}

\subsection{Mean-Field framework}
\label{mft}

It is well known that in the limit $N \to \infty$, the contribution of the interaction term in Eq.~\eqref{eq:ge_y} can be represented as a Gaussian noise, $\eta_i = \lim_{N\to\infty}\sum_{j=1}^N J_{ij} F(y_j)$ where each element of the connectivity matrix $J$ is drawn independently from a Gaussian distribution with zero mean and variance $g^2/N$. Here, the parameter $g$ controls the overall strength of network interactions. Consequently, the dynamics of the expression level of the $i^{th}$ agent reduces to
\begin{align*}
    \dot{y}_i
    &= \eta_i - y_i + \theta_i +\dot{\theta}_i + c_i.
\end{align*}

From this point onward, we aim to handle the above equations mathematically to construct a DMFT framework that incorporates the slow feedback term. Originally, this term appears as a feedback in the equations, but under an appropriate transformation, it can be treated within a single effective equation, where the feedback influences the transformed dynamics along with its time derivative.

Following the method introduced by Sompolinsky, many studies have applied DMFT to Hopfield-type or Hopfield-like models, often including periodic driving terms, to analyse randomness, chaos, and potential landscape deformation in network dynamics~\cite{Sompolinsky1988,crisanti1987dynamics,rajan2010stimulus,panda2017chaos,sompolinsky1982relaxational,engelken2023lyapunov}. Such models are relatively easier to analyse using DMFT because the external periodic term introduces regular, time-dependent perturbations~\cite{Sompolinsky1988,crisanti1987dynamics,rajan2010stimulus,panda2017chaos,sompolinsky1982relaxational}.

However, there remains a gap in understanding models with slowly varying, time-dependent terms, such as the theta-dependent feedback in our equations. Unlike periodic driving, the slow epigenetic feedback cannot be approximated as a simple oscillatory term, making its treatment within conventional DMFT challenging.

This raises the natural question: how can DMFT be applied to this model, and how does the transformed local dynamics--and hence the global dynamics--reshape the potential landscape? Constructing such a landscape would provide insight into cellular differentiation under slow epigenetic regulation, which gradually reprograms gene-expression dynamics during development~\cite{matsushita2022dynamical,kaneko2007evolution,miyamoto2015pluripotency,KanekoPRR2020}.

\subsection{Disparity of characteristic timescales}
\label{disparity}

Under the assumption $\nu \ll 1$, we observe a clear separation of timescales for $x_i$ and $\theta_i$. The remaining parameters are unaffected in this limit. As a consequence, we have $\dot{\theta}_i \approx 0$ when deriving the equation for $\dot{y}_i$.
Importantly, this does not mean that $\theta_i$ is truly time-independent: 
the final solution for $y_i$ still depends on the full, slowly varying $\theta_i$. 
Depending on the timescale of interest, one can therefore either treat $\theta_i$ as effectively constant (fast dynamics of $x_i$) 
or account for its slow temporal evolution (longer timescales).
Which approximation is appropriate should be clear from the context, based on the current time-scale of interest.
Under this assumption, the governing equation for $y_i$ of the $i^{\text{th}}$ agent becomes
\begin{align}
\label{eq:y_system}
    \dot{y}_i &= \eta_i - y_i + \theta_i + c_i .
\end{align}

We further treat this equation by separating the regular and noisy contributions via the decomposition $y_i = y_{i,0} + y_{i,1}$, resulting in
\begin{align}
    \label{eq:atomic_0}
    \dot{y}_{i,0} &= -y_{i,0} + \theta_i + c_i ,\\
    \label{eq:atomic_1}
    \dot{y}_{i,1} &= -y_{i,1} + \eta_i .
\end{align}

Solving these ODEs yields the expressions
\begin{align*}
    y_{i,0} &= \underbrace{\left(\theta_i + c_i\right)}_{\tilde{A}_i}\!\left[1 - \exp\left(-t\right)\right] + \tilde{y}_i \exp(-t) , \\
    y_{i,1} &= \exp(-t) \int_0^t \eta_i(t') \exp(t') \,\mathrm{d}t' ,
\end{align*}
where $\tilde{y}_i = y_i(t=0)$ is the initial condition.

More formally, we identify three characteristic timescales associated with significant changes in the system dynamics: $\tau_0 = 1$ for $y_i$, $\tau_1 = 1/\nu \gg \tau_0$ for $\theta_i$, and an intermediate timescale $\tau_0 \ll T \ll \tau_1$ that remains compatible with this separation.

This timescale separation has an important impact on the evolution of $\theta_i$. In particular, the contribution of $x_i(t)$ in Eq.~\eqref{eq:ge_theta} can be replaced by an effective, time-independent averaged field over the intermediate window $T$:
\begin{align}
    X_i(\theta) = \frac{1}{T} \int_0^T x_i(t)\,\mathrm{d}t.
    \label{eq:xdynamics}
\end{align}

This averaged quantity still depends on $\theta_i$, and leads to the following effective governing equation for $\theta_i$:
\begin{align}
    \dot{\theta}_i &= \nu \left[ \alpha\, X_i(\theta) - \theta_i \right].
    \label{eq:thetadynamics}
\end{align}

The derivation of the averaged field $X_i$ constitutes the main technical challenge and will be carried out in the following sections.

\section{Deriving the corresponding average-field approximation}
\label{averagefield}

For further analysis toward achieving the goal of constructing the autocorrelation function (its importance and derivation are described in Sec.~\ref{autocorelation}), we consider the dynamics arising from Eqs.~\eqref{eq:xdynamics} and \eqref{eq:thetadynamics}. The resulting autocorrelation function is expected to depend solely on the slowly varying, feedback-driven term $\theta$. To this end, special care must be taken in treating these equations, since the original dynamics and the dynamics obtained after separating the time scales via the averaged-field approach remain coupled. This coupling persists even after the time-scale separation and therefore requires careful investigation. The detailed analysis addressing this issue is presented below. 

\subsection{Representation $X_i$ in terms of $y_i$}
\label{representation}

First, we express the original variable $x_i$ in terms of the reduced variable $y_i$. We define

\begin{align*}
    X_i
    &= \frac{1}{T} \int_0^T x_i(t) \mathrm{d} t  \\
    &= \frac{1}{T} \int_0^T \{F\left[y_i(t)\right] - \dot{x}_i(t)\} \mathrm{d} t
    \\
    &= \frac{1}{T} \int_0^T F\left[y_i(t)\right] \mathrm{d} t- 
    \underbrace{\frac{1}{T} \left[x_i(T) -x_i(0)\right]}_{\approx 0} \\
    &= \frac{1}{T} \int_0^T \tanh\left[ \beta \, y_i(t)\right] \mathrm{d} t.
\end{align*}

We can neglect the second term because $x_i(t)$ is bounded within $[-1,1]$ for all $t$ (assuming the initial condition satisfies $x_i(0)\in[-1,1]$), so the difference $x_i(T)-x_i(0)$ is at most $\pm 2$. Since $T \gg \tau_0 = 1$, the contribution of this term vanishes in the limit of large $T$.

\subsection{Evaluating the expected noise contribution}
\label{noise}

In the presence of non-negligible noise, we must evaluate the contribution arising from the noise term. To do so, we generalize $X_i$ so that it is averaged not only over time $t$ but also over the noise realization $\eta$, which itself is time-dependent. This can be interpreted as treating the coupling of a given agent to the rest of the system, and hence as accounting for the influence of the agent’s surroundings.  In this context, we make use of the complex integral relation
\begin{align}
    \label{eq:relation_tanh}
    \tanh(z)
    &=
    -\frac{2 i}{\pi}\int_0^\infty
    \frac{t^{\frac{2iz}{\pi}}-1}{t^2-1} \mathrm{d}t,
\end{align}
with $-\pi/2<\text{Im}(z)<0$. Thus, in order to apply this relation, we introduce a small complex parameter $\epsilon$, which leads to
\begin{align}
    X_j 
    &=
    \mathbb{E}_{\eta, t}\big[\tanh\left\{\beta\, y_j\left[t, \eta(t)\right]\right\}\big] \nonumber \\
    &=
    \lim_{\epsilon\rightarrow 0^+}
    \mathbb{E}_{\eta, t} \left[ \tanh\left\{ \beta
    \left[ y_j^0(t)+ y_j^1(t, \eta(t))\right]
    - i \epsilon
    \right\}\right] \nonumber \\
    &=
    \lim_{\epsilon\rightarrow 0^+}
    \mathbb{E}_{\eta, t} \left[
    \frac{-2 i}{\pi}\int_0^\infty
    \frac{u^{\frac{2i}{\pi}\left(\beta
    \left[ y_j^0(t)+ y_j^1(t, \eta(t))\right]
    - i \epsilon\right)}-1}{u^2-1} \mathrm{d}u
    \right] \nonumber \\
    &=
    \label{eq:X_average_field}
    \lim_{\epsilon\rightarrow 0^+}
    \frac{-2i}{\pi}\mathbb{E}_{t}\left[
    \int_0^\infty
    \left(
    u^{\frac{2i}{\pi}\beta
    \left[y_j^0(t) - i \epsilon\right]}
    H(u,t)
    -1
    \right)
\frac{\mathrm{d}u}{u^2-1}
    \right]
\end{align}
where we have used the exponential identity $x^a = \exp\left[a \log(x)\right]$ and defined
\begin{align}
\label{eq:noise_expectation}
    H(u,t) = \mathbb{E}_{\eta} \left\{
    \exp \left[\frac{2i\beta
    }{\pi} y_j^1(t, \eta(t)) \log(u)\right]
    \right\}.  
\end{align}
Notice that for Gaussian noise with zero mean and bounded, possibly time-dependent variance, 
the noise average produces a bounded analytic contribution. 
This allows us to perform the noise and time averages on the regularized integrand at fixed $\epsilon>0$.
The exchange of the noise and time averages with the $u$-integral is physically justified by the damping provided by the noise average.

Let us first focus on the newly defined expectation over the noise within the above integrals. By incorporating our previous assumption that $\eta$ represents a Gaussian noise as well as using Isserlis Theorem~\cite{isserlis1918}, we can treat the noise expectation. We obtain the following-now index independent-relation
\begin{align}
   \label{eq:treated_noise_expectation} H(u,t)=\exp\left(-\frac{4 \beta^2 g^2}{\pi^2} \tilde{C}(t)\right),
\end{align}
which is depending on the effective autocorrelation function $\tilde{C}(t)$, where, 
\begin{align*}
    \tilde{C}(t)=\int_0^t\exp(\tau-2t)\left[\int_0^\tau C_{\eta}(s) \mathrm{d}s\right]\mathrm{d}\tau.
\end{align*}

The above relation reveals a subtle issue related to the effective autocorrelation function, as it depends explicitly on the autocorrelation function $C_{\eta}(t)$ of the underlying Gaussian process $\eta$. A detailed derivation of the Eq. \eqref{eq:treated_noise_expectation} can be found in Appendix \ref{app:treating_noise_expectation}.

\subsection{Evaluation of the $t-$integral}
With the new form of the noise term, we aim to perform the time integration in order to simplify the subsequent $u$-integration. The current expression for the averaged field is given in Eq.~\eqref{eq:X_average_field}, which contains a single time-dependent contribution. For the moment, we restrict ourselves to the reduced integral
\begin{align*}
     I_1
     &= 
    \frac{1}{T}
    \int_0^T  
   u^{\frac{2i}{\pi}\beta
    \left[y_j^0(t) - i \epsilon\right]}
    \exp\left[
    -\frac{4 \beta^2 g^2}{\pi^2}
    \tilde{C}(t)
    \log^2 (u)
    \right]
    \mathrm{d}t
    \\
    &= 
    \frac{1}{T}
    \int_0^T  
    e^{i a y_j^0(t) + c}
    e^{- b\tilde{C}(t)}
    \mathrm{d}t   
\end{align*}
with the $u$-dependent parameters,
\begin{align*}
    a
    &= 
    \frac{2\beta}{\pi} \log(u),\\
    b
    &=
    \frac{4 \beta^2 g^2}{\pi^2}
    \log^2(u),\\
    c
    &=
    \frac{2\beta}{\pi}\epsilon \log(u).
\end{align*}

We substitute the expressions for $y(t)$ and apply the series expansion of the exponential  leading to the following derivation
\begin{align}
\label{eq:full_time_integral}
     I_1
    &= 
    \frac{1}{T}
    \int_0^T  
    e^{i a y_j^0(t) + c}
    e^{- b\tilde{C}(t)}
    \mathrm{d}t \nonumber \\
    &=
    \frac{1}{T}
    \int_0^T  
    \exp\left[i a \left(
    \tilde{A}_j
    +(\tilde{y}_j-\tilde{A}_j)e^{-t}
    \right) + c\right]
    \nonumber \\
    &\hspace{0.5cm}\times
    \exp\left(- b\,
    \tilde{C}(t)
    \right)
    \mathrm{d}t
    \nonumber \\
    &=
    e^{ia\tilde{A}_j+c}\sum_{n,k}\frac{(ia)^n (-b)^k}{n! k!}
    (\tilde{y}_j-\tilde{A}_j)^n 
    \nonumber \\
    &\hspace{0.5cm}\times
    \frac{1}{T}
    \int_0^T  
    e^{-nt}
    \tilde{C}^k(t) \,
    \mathrm{d}t.
\end{align}

The above equation managed to reformulate the original time-integration to an infinite sum over polynomial contributions within a new integral.
Thus, let us now focus specifically on the new integral involving the effective autocorrelation function for the case $n \neq 0$ or $k \neq 0$:
\begin{align}
    I_2
    &=
    \frac{1}{T}
    \int_0^T  
    e^{-nt}
    \tilde{C}^k(t) \,
    \mathrm{d}t
    \nonumber
    \\
    &=
    \int_0^T  
    \frac{e^{-nt}}{T}
    \left\{
    \int_0^t e^{\tau - 2 t} 
    \left[ \int_0^\tau C(\tau') \mathrm{d}\tau' \right]
    \mathrm{d}\tau
    \right\}^k
    \mathrm{d}t  \nonumber \\
    &=
    \int_0^T  
    \frac{e^{-(n+2k)t}}{T}
    \left\{
    \int_0^t e^{\tau}
    \left[ \int_0^\tau C(\tau') \mathrm{d}\tau' \right]
    \mathrm{d}\tau
    \right\}^k
    \mathrm{d}t  \nonumber \\
    &\leq
    \label{eq:relation_C}
    \frac{1}{T}
    C(0)^k
    \int_0^T  
    e^{-(n+2k)t}
    \left[
    (t-1)e^t -1
    \right]^k
    \mathrm{d}t  \nonumber \\
    &\leq
    \frac{1}{T}
    C(0)^k
    \int_0^T  
    e^{-(n+2k)t}
    t^k
    e^{kt}
    \mathrm{d}t  \nonumber \\
    &=
    \frac{1}{T}
    \frac{C(0)^k}{(n+k)^{k+1}}
    \int_0^\frac{T}{n+k}  
    e^{-t'}
    \,
    t'^k
    \,
    \mathrm{d}t' \\
    &\leq
    \label{eq:incomplete_gamma_relation}
    \frac{1}{T}
    \frac{C(0)^k}{(n+k)^{k+1}}
    \Gamma\left(k+1\right) \nonumber \\
    &=
    \frac{1}{T}
    \frac{k!}{(n+k)^{k+1}} C(0)^k.
\end{align}

We used the bound $|C(\tau')| \leq C(0)$ in Eq.~\eqref{eq:relation_C}, together with the fact that the incomplete Gamma function $\Gamma(s,x)$ is upper bounded by the complete Gamma function $\Gamma(s)$ in Eq.~\eqref{eq:incomplete_gamma_relation}. From this, we see that the above integral scales as $1/T$ as long as $n \neq 0$ or $k \neq 0$. Furthermore, for the case $n = k = 0$, the integrand becomes trivial, yielding the full solution,
\begin{align*}
    I_2
    &\leq
    \begin{cases}
    1
    &\text{,     for $n=k=0$}\\
    \frac{1}{T}
    \frac{k!}{(n+k)^{k+1}} C(0)^k
    &\text{,     else,}
    \end{cases}
    \nonumber \\
    &\approx
    \delta_{n,0} \delta_{k,0},
\end{align*}
where the approximation is valid for $T \gg \tau_0 = 1$. Therefore, we obtain the result for the full integral in Eq.~\eqref{eq:full_time_integral}, given by
\begin{align}
    I_1
    &\approx
    e^{ia\tilde{A}_j+c} \sum_{n,k} \frac{(ia)^n (-b)^k}{n! k!}
    (\tilde{y}_j-\tilde{A}_j)^n 
    \delta_{n,0}
    \delta_{k,0} \nonumber \\
    &= 
    e^{ia\tilde{A}_j+c}
\end{align}

In addition, this gives us the reduced form of the full integral as
\begin{align*}
     X_j(t)
    \approx
    \lim_{\epsilon\rightarrow 0^+}
    \frac{-2i}{\pi} 
    \int_0^\infty
    \frac{
    u^{\frac{2i\beta}{\pi}\tilde{A}_j}
    u^{\frac{2\beta}{\pi}\epsilon}
    -1}{u^2-1}
    \mathrm{d}u.
\end{align*}
This derivation demonstrates that the result does not depend on the specific form of $C(t)$, as long as the noise remains Gaussian.

\subsection{Handling the u-integral}
We can now reformulate our result by applying the previous integral relation \eqref{eq:relation_tanh} for the hyperbolic tangent, yielding the following:
\begin{align*}
    X_j(t)
    &=    \lim_{\epsilon\rightarrow 0^+}
    \frac{-2i}{\pi} 
    \int_0^\infty
    \frac{1}{u^2-1}
    \left\{
    u^{\frac{2i\beta}{\pi}\tilde{A}_j}
    u^{\frac{2\beta}{\pi}\epsilon}
    -1
    \right\}
    \mathrm{d}u  \\
    &=    \lim_{\epsilon\rightarrow 0^+}
    \frac{-2i}{\pi} 
    \int_0^\infty
    \frac{
    u^{\frac{2i}{\pi}\left(\beta\tilde{A}_j-i\epsilon\right)
    }
    -1
    }{u^2-1}
    \mathrm{d}u \\
    &=
    \tanh(\beta \tilde{A}_j)  \\
    &=
    \tanh\left[\beta \left(\theta_j + c_j\right)\right]
\end{align*}
Thus, the average field of the $i^{th}$ agent only depends on the response strength $\beta$, the external input $c_i$ as well as the instantaneous threshold $\theta_i(t)$. 

\section{Analytical Approach to Solving $\theta$}
To solve the slow dynamics of $\theta_i$ arising from Eq.~\eqref{eq:thetadynamics} using the above averaged field approximation, we have
\begin{align}
\label{eq:theta_final}
    \dot{\theta}_i &= \nu\left(\alpha X_i - \theta_i\right)  \nonumber \\
    &= \nu\alpha \tanh\left[\beta(c_i+\theta_i)\right] - \nu \theta_i 
\end{align}

This equation can be approached via separation of variables:
\begin{align*}
    \frac{\mathrm{d}\theta_i}{\mathrm{d}t}
    &= \nu\alpha \tanh\left[\beta(c_i+\theta_i)\right] - \nu \theta_i 
     \nonumber \\
    \nu \int\mathrm{d}t 
    &=
    \int\frac{1}{\alpha\tanh\left[\beta(c_i+\theta_i)\right]-\theta_i}\mathrm{d}\theta_i \nonumber \\
    &=
    \int\frac{1}{\alpha\beta\tanh\left(x\right)-x+\beta c_i}\mathrm{d}x.
\end{align*}

However, this integral has no known analytic solution and likely does not admit one.

\section{Autocorrelation Function and Time-Scale Separation in the DMFT Framework}\label{autocorelation}
The autocorrelation function is the central object in DMFT, as originally introduced in Sompolinsky’s seminal work~\cite{Sompolinsky1988,crisanti1987dynamics,panda2017chaos}, where it encapsulates the reduction of complex global network dynamics to an effective local stochastic process through the matching of moments, including higher-order correlations~\cite{Sompolinsky1988,crisanti1987dynamics,panda2017chaos,engelken2023lyapunov}. This framework enables the study of large random networks by mapping their collective dynamics onto an effective single-degree-of-freedom description~\cite{Smita,Sompolinsky1988,crisanti1987dynamics,panda2017chaos}. Within this perspective, the resulting effective dynamics can be interpreted through the lens of Newtonian mechanics, where the complexity of the network is absorbed into an effective potential and noise term, providing insight into the global behaviour of the system~\cite{Sompolinsky1988,crisanti1987dynamics,panda2017chaos}.

A natural question then arises: to what extent are the network dynamics modified by the presence of a slow epigenetic modification term in a GRN, given that the prototype model without such a term has been extensively studied, beginning with Sompolinsky’s original work~\cite{Sompolinsky1988}. In particular, can the autocorrelation function still capture the essential effects of epigenetic reprogramming when slow epigenetic variables modulate the regulatory interactions? Furthermore, how is the effective Newtonian dynamics altered through changes in the autocorrelation function induced by epigenetic modifications? Specifically, how does the corresponding effective potential deform in the presence of such slow epigenetic feedback, and to what extent does it reshape the dynamical landscape of the system?

Since DMFT becomes exact only in the limit of infinitely large networks, finite-size effects inevitably introduce deviations from the mean-field description. To address this, we restrict our analysis to the finite-size N-particle autocorrelation function of the transformed system and examine how slow epigenetic modifications manifest within this framework. Starting with
\begin{align}
C_N(\tau) &= \frac{1}{N} \sum_{i=1}^N \frac{1}{T} \int_0^T \mathrm{d}t \, \mathbf{E}_J \left\{ F[y_i(t)] F[y_i(t+\tau)]\right\} \nonumber\\
\label{eq:emperical_auto_correlation}
&=\frac{1}{N}\sum_{i=1}^N \mathbf{E}_{t,J} \left\{ F[y_i(t)] F[y_i(t+\tau)] \right\}.
\end{align}

We begin by deriving a relationship between the noise autocorrelation function and the expression above. Imposing the self-consistency condition on the couplings, we find that the effective noise must satisfy the following relations:
\begin{align*}
   &\mathbf{E}_{t,\eta} \left[\eta_i(t)\right] = 0,\\
   &\mathbf{E}_{t,\eta} \left[\eta_i(t) \eta_j(t+\tau)\right]
    =
    \delta_{ij} g^2 C(\tau),
\end{align*}
with $ C(\tau) \equiv C_\infty(\tau)$.
A more detailed derivation is provided in Appendix~\ref{app:self_consistency}.

In the next step, we define the atomic autocorrelation function $\Delta(\tau) = \mathbf{E}_{t,J} \left[ y_{k,1}(t) y_{k,1}(t+\tau) \right]$. By substituting Eq.~\eqref{eq:atomic_1} into this definition, we obtain the following differential equation,
\begin{align}
    \label{eq:pde_autocorrelation}
    \frac{\mathrm{d}^2 \Delta(\tau)}{\mathrm{d}\tau^2} = \Delta(\tau) - g^2 C(\tau).
\end{align}

Note that the integration is performed over the finite interval \( [0, T] \), rather than extending to infinity. Consequently, the averaging is restricted to the relevant short-time scale. As a result, an implicit dependence on the longer timescales is retained in the dynamics.

By recognizing that \( y_{k,1}(t) \) is driven by Gaussian noise---and is therefore not specific to the \( k \)-th agent---we may drop the index \( k \). Since these variables are noise-driven, they are themselves Gaussian random variables, fully characterized by their moments,
\begin{align*}
    & \mathbf{E}_{t,J} \left[ y_1(t) \right] = \mathbf{E}_{t,J} \left[ y_1(t+\tau) \right] = 0,    \\
    & \mathbf{E}_{t,J} \left[ y_1(t)  y_1(t+\tau) \right] = \Delta(\tau).
\end{align*}

These constraints can be readily obtained by introducing three independent Gaussian random variables \( z_1, z_2, z_3 \) with zero mean and unit variance, which are related to the original random variable through the following transformation:
\begin{align*}
    y_1(t) &= \hat{\alpha}(\tau) z_1 + \hat{\beta}(\tau) z_3, \\
    y_1(t+\tau) &= \hat{\alpha}(\tau) z_2 + \hat{\gamma}(\tau) z_3,
\end{align*}
with $\hat{\alpha}(\tau) = \sqrt{\Delta(0) - |\Delta(\tau)|}$, $\hat{\beta}(\tau)= \mathrm{sgn}\left[\Delta(\tau)\right] \sqrt{|\Delta(\tau)|}$ and $\hat{\gamma}(\tau)=\sqrt{|\Delta(\tau)|}$. We can now express the autocorrelation function \( C(\tau) \) as
\begin{align}
\label{eq:autocorrelation}
  C(\tau) =&\lim_{N\to\infty} \mathbf{E}_{N} \Bigg\{\int_{-\infty}^\infty \int_{-\infty}^\infty \int_{-\infty}^\infty \mathrm{D}z_3 \, \mathrm{D}z_2 \, \mathrm{D}z_1 \, 
  \nonumber \\
  &\times 
  F\left[\hat{\alpha}(\tau) z_1 + \hat{\beta}(\tau) z_3 + (\theta_i^t + c_i)\right] \nonumber \\
  &\times 
  F\left[\hat{\alpha}(\tau) z_2 + \hat{\gamma}(\tau) z_3 + (\theta_i^{t+\tau} + c_i)\right]\Bigg\},
\end{align}
where we used $y_i(t) = y_{i,0}(t)+y_{i,1}(t)\equiv y_{i,0}(t)+y_{1}(t)$ and $\mathrm{D}z_j = \frac{1}{\sqrt{2\pi}} \exp(-z_j^2/2) \mathrm{d} z_j$. 

Furthermore, we assume that the system is in a quasi-steady state and follows the reduced dynamics adiabatically, as dictated by \(\theta\). This allows us to eliminate the dependence on initial conditions in \(C(t)\). Although the time dependence of \(\theta\) does not significantly affect the time-integral over the short interval, it does influence the dynamics for \(t \gg T\). As a result, \(C(\tau)\) retains an implicit dependence on the long-time variable \(t\). To reflect this, we adopt the notation \(C^t(\tau)\), where the superscript \(t\) denotes this slow-time dependence.


Since we aim to solve Eq.~\eqref{eq:pde_autocorrelation}, we note that the timescales associated with \(\theta\) remain much longer than those relevant to the ODE, which are determined by the intrinsic time scale \(\tau_0\) of \(y\). Consequently, the equation can be further simplified, yielding the following expression: 
\begin{align*}
  C^t(\tau) = &\lim_{N\to\infty} \mathbf{E}_{N} \Bigg\{\int_{-\infty}^\infty \int_{-\infty}^\infty \int_{-\infty}^\infty \mathrm{D}z_3 \, \mathrm{D}z_2 \, \mathrm{D}z_1 \,
  \\
  &\times
  F\left\{\hat{\alpha}(\tau) z_1 + \hat{\beta}(\tau) z_3 + \left[\theta_i^t + c_i\right]
  \right\} \\
  &\times
  F\left\{\hat{\alpha}(\tau) z_2 + \hat{\gamma}(\tau) z_3 + \left[\theta_i^{t} + c_i\right]\right\}\Bigg\}.
\end{align*}

This simplification allows us to write Eq. \eqref{eq:pde_autocorrelation} through a potential as, 
\begin{align}
\label{eq:pde_autocorrelation_potential}
    \frac{\mathrm{d}^2\Delta(\tau)}{\mathrm{d}\tau^2} &= - \frac{\mathrm{d}V^t(\Delta)}{\mathrm{d}\Delta}.
\end{align}

By slightly redefining the parameters--factoring \(g\) out of the variance and incorporating it into the gain parameter \(\beta\)--we obtain the dimensionless effective gain \(\beta g\). This rescaling does not alter the system dynamics but simplifies the subsequent derivations. Accordingly, we can define an effective potential \(V^t(\Delta)\) of the form:
\begin{align}
\label{eq:effective_potential}
    V^t(\Delta)
    &=
    -\frac{\Delta^2}{2}+ 
    \lim_{N\to\infty} \mathbf{E}_{N} \Bigg\{
    \int_{-\infty}^{\infty}
    \int_{-\infty}^{\infty}
    \int_{-\infty}^{\infty}
    \mathrm{D}z_3
    \mathrm{D}z_2
    \mathrm{D}z_1
    \nonumber \\
    &\times
    \Phi\left[ \sqrt{\Delta_0 - |\Delta|} z_2 + \sqrt{|\Delta|}z_3 + (\theta_i^t + c_i)\right] \nonumber \\
    &\times
    \Phi\left[ \sqrt{\Delta_0 - |\Delta|} z_1 + \sqrt{|\Delta|}z_3 + \mathrm{sgn}(\Delta) (\theta_i^t + c_i)\right]\Bigg\},
\end{align}
for all $\Delta$. We further use the following novel function  
\begin{align*}
    \Phi(x) = \frac{1}{\beta g} \ln{\cosh{\beta g x}}.
\end{align*}

A detailed proof that the above potential indeed satisfies Eq.~\eqref{eq:pde_autocorrelation_potential} is provided in Appendix~\ref{app:proof_potential}.

\subsection{Viable Autocorrelation States}\label{sec:auto_regimes}

The system's autocorrelation function $\Delta$ is subject to several constraints: boundedness ($\Delta_0 \geq |\Delta|$), evenness ($\Delta(\tau) = \Delta(-\tau)$), and no initial momentum ($\dot{\Delta}(\tau=0) = 0$). Consequently, the following regimes are discussed only for $-\Delta_0 \leq \Delta \leq \Delta_0$, which defines the most rudimentary valid range of $\Delta$. For $\theta = 0 = c$, we recover the same potential and derivations as in Sompolinsky's work. In this section, we focus solely on the viability of certain autocorrelation functions. The stability of these states--while related to the potential--cannot be directly inferred from its quantitative behaviour and will be discussed separately.  

To illustrate the possible regimes, we initially consider the simplified scenario where all $N$ agents share the same parameters, $\theta_i \equiv \theta$ and $c_i \equiv 0$. This simplification aids reproducibility and clarity, though the general features of the potentials persist even when these restrictions are relaxed. By plotting potential \eqref{eq:effective_potential} for various parameter choices, we can distinguish different states the system may occupy.

\begin{figure}[t]
    \includegraphics[width=\linewidth]{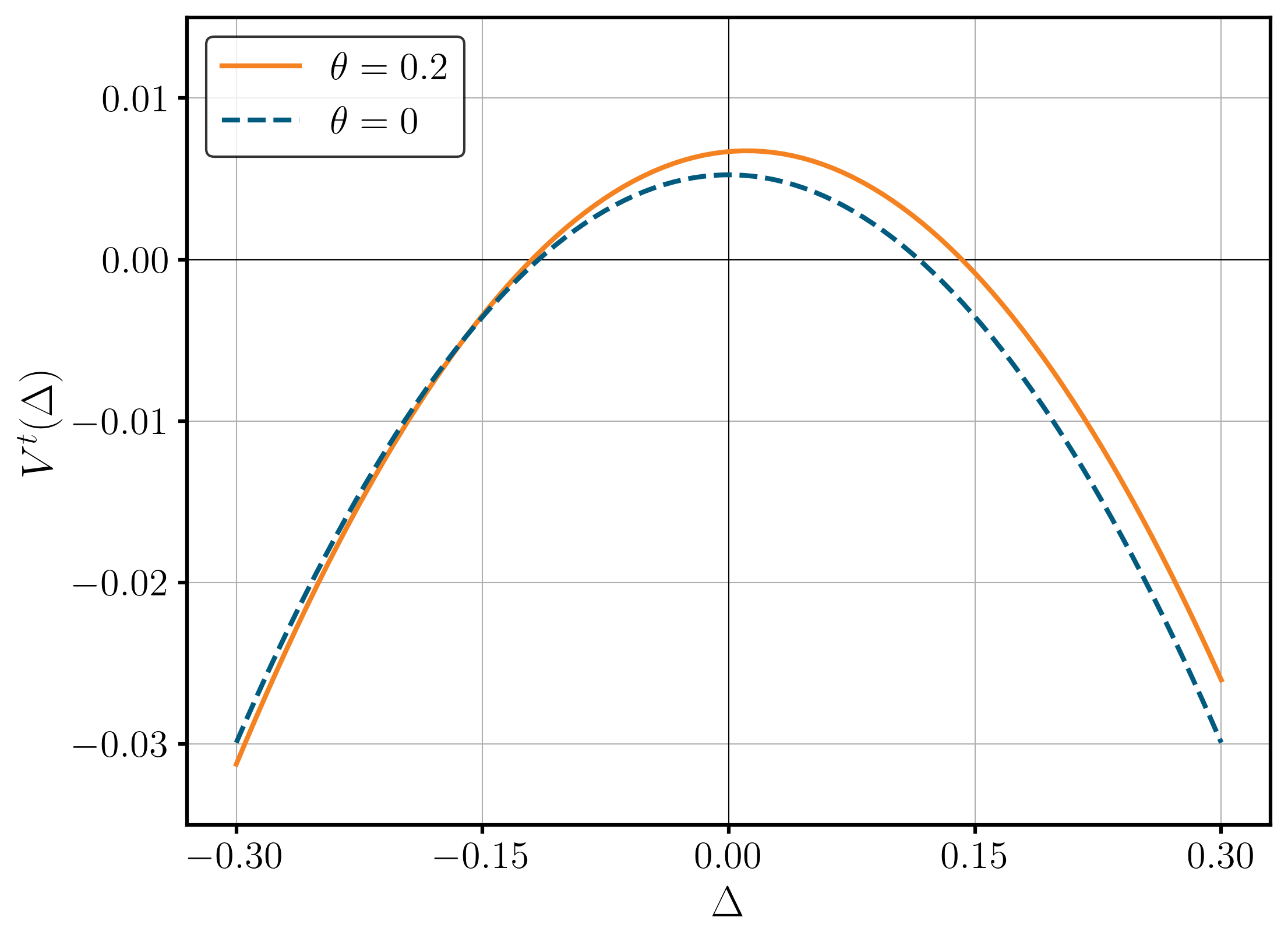}
    \caption{Potentials for Regime 1 for different values of $\theta$.}
    \label{fig:potential_regime_1}
\end{figure}

\begin{figure}[t]
    \includegraphics[width=\linewidth]{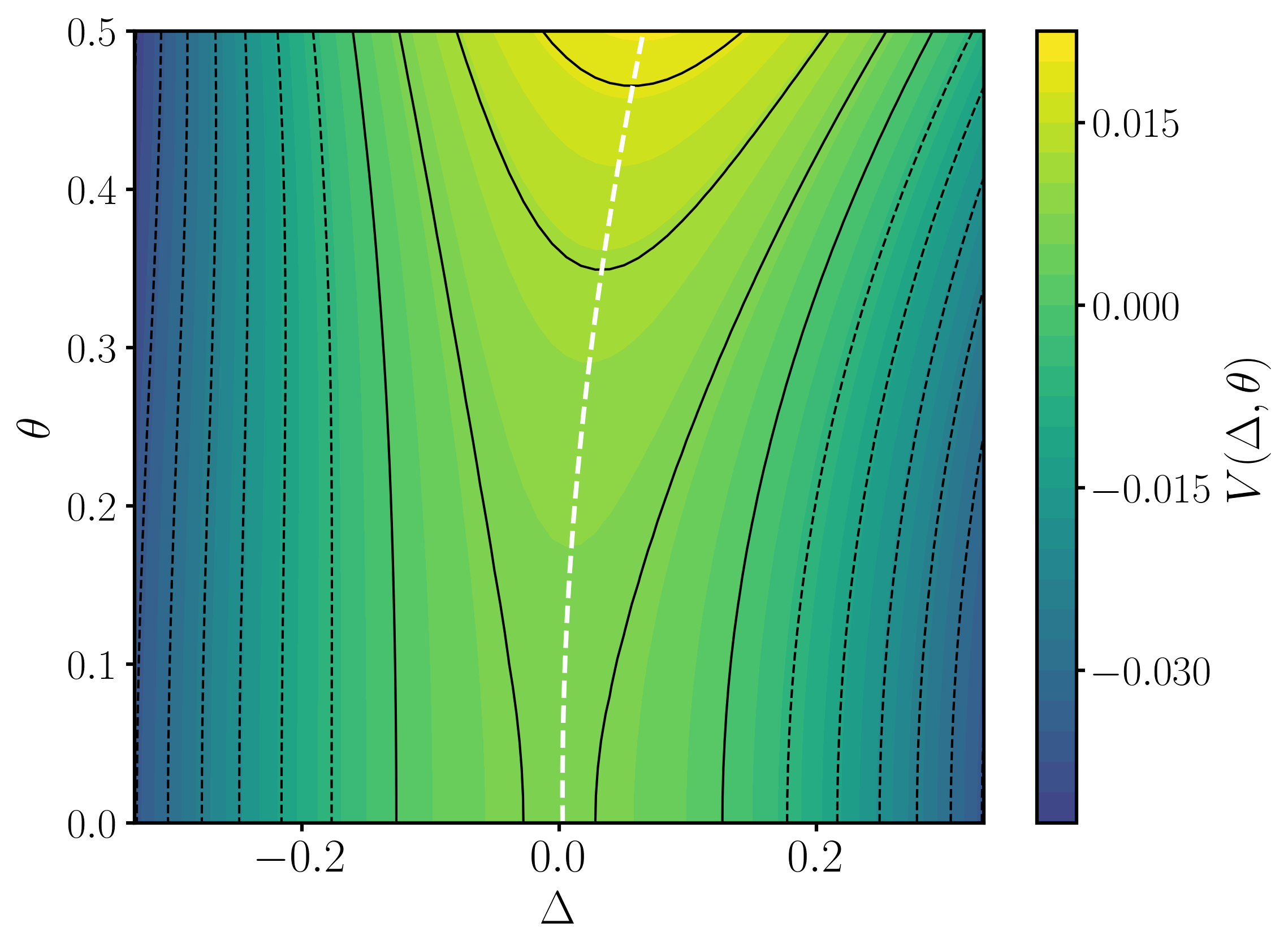}
    \caption{Dependence of Regime 1 on different values of $\theta$, with $c=0$. The white dashed line indicates the location of the central maximum for each value of $\theta$.}
    \label{fig:potential_contour1}
\end{figure}

The \textit{first regime}, shown in Fig. \ref{fig:potential_regime_1}, occurs for dimensionless gain parameters $\beta g < 1$. The system admits only one classically viable solution satisfying $\dot{\Delta}(0) = 0$, namely the trivial solution $\Delta(t) = 0$ for $\theta = 0$. This indicates that the system described by Eq.~\eqref{eq:y_system} flows toward a trivial fixed point where the activity tends to zero. Allowing $\theta \neq 0$ introduces slight shifts in the position of the minimum, which can be interpreted as the system settling into a non-zero fixed point or, equivalently, a spin glass freezing. This shift is also visible on the white line of the density plot in Fig. \ref{fig:potential_contour1}, where the curvature shows that larger values of $\theta$ move the central maximum to the right.

\begin{figure}[t]
    \includegraphics[width=\linewidth]{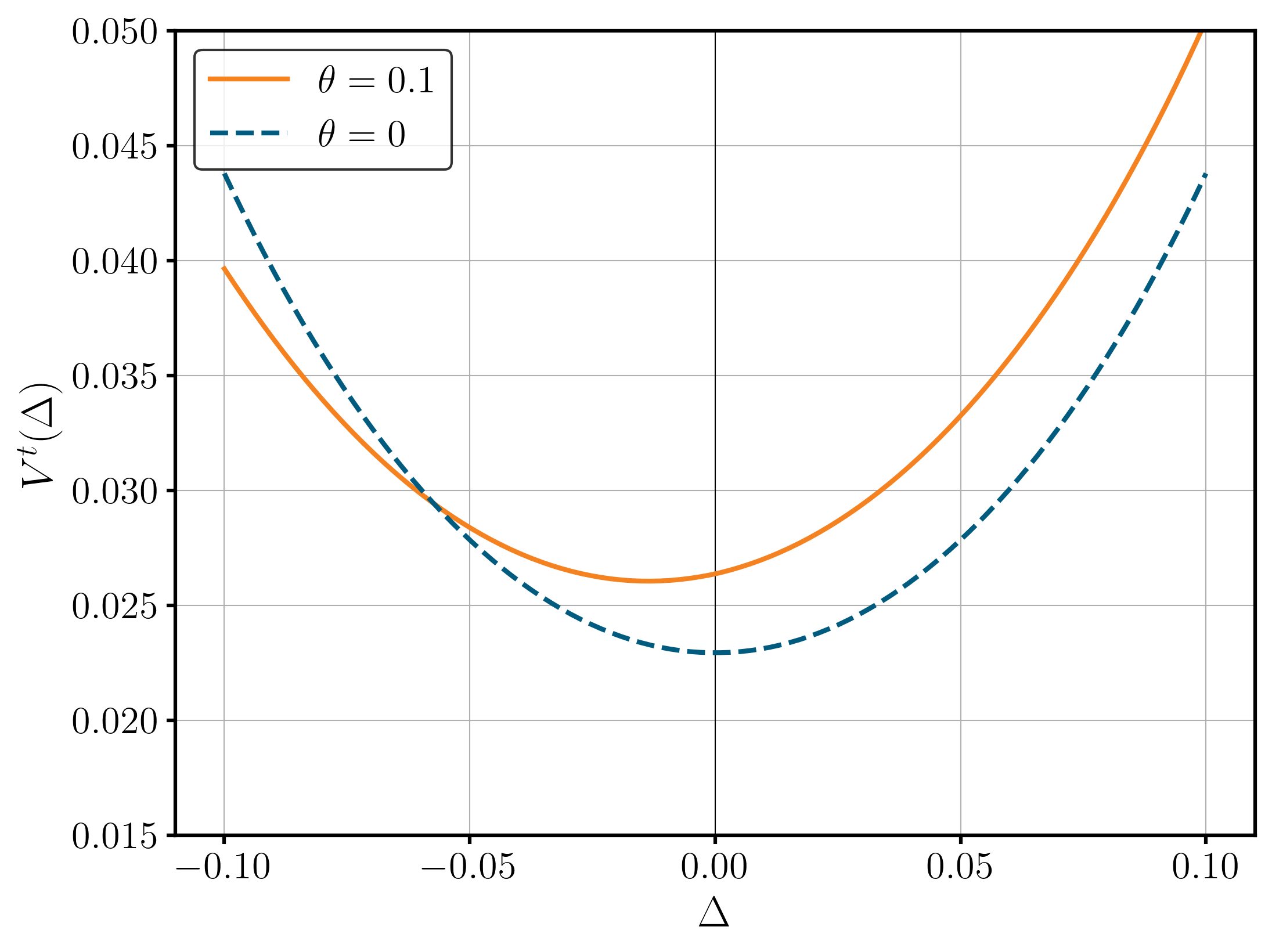}
    \caption{Potentials for Regime 2 for different values of $\theta$.}
    \label{fig:potential_regime_2}
\end{figure}

The \textit{second regime} arises for $\beta g > 1$ and $\Delta_0 < \tilde{\Delta}_1$ for some specific value $\tilde{\Delta}_1$, as illustrated in Fig. \ref{fig:potential_regime_2}. For a given $V(\Delta)$ with $\theta = 0$, there exists a continuum of self-consistent solutions of Eq. \eqref{eq:pde_autocorrelation}, corresponding to periodic orbits and limited cycles. Non-zero values of $\theta$ distort and shift the potential, introducing asymmetry. This asymmetry eliminates some previously allowed limiting cycles, since they would violate the boundedness constraint ($\Delta > \Delta_0$). The only remaining allowed solution is the potential minimum, which represents a non-zero fixed point of the system--a spin glass freezing state. The location of this fixed point can be tuned by adjusting $\theta$, offering a way to control the system’s behavior.

\begin{figure}[t]
    \includegraphics[width=\linewidth]{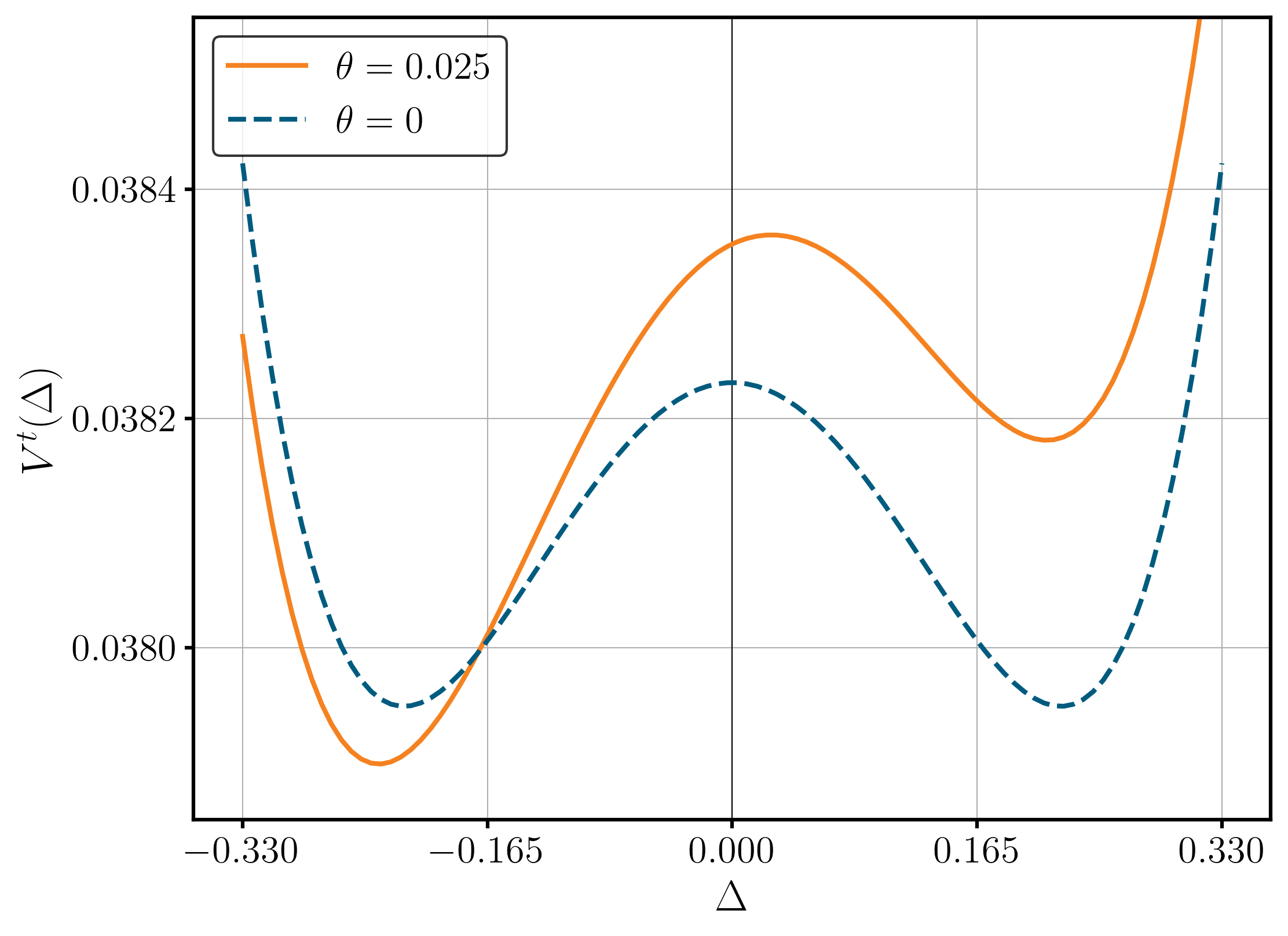}
    \caption{Potentials for Regime 3 for different values of $\theta$.}
    \label{fig:potential_regime_3}
\end{figure}

\begin{figure}[t]
    \includegraphics[width=\linewidth]{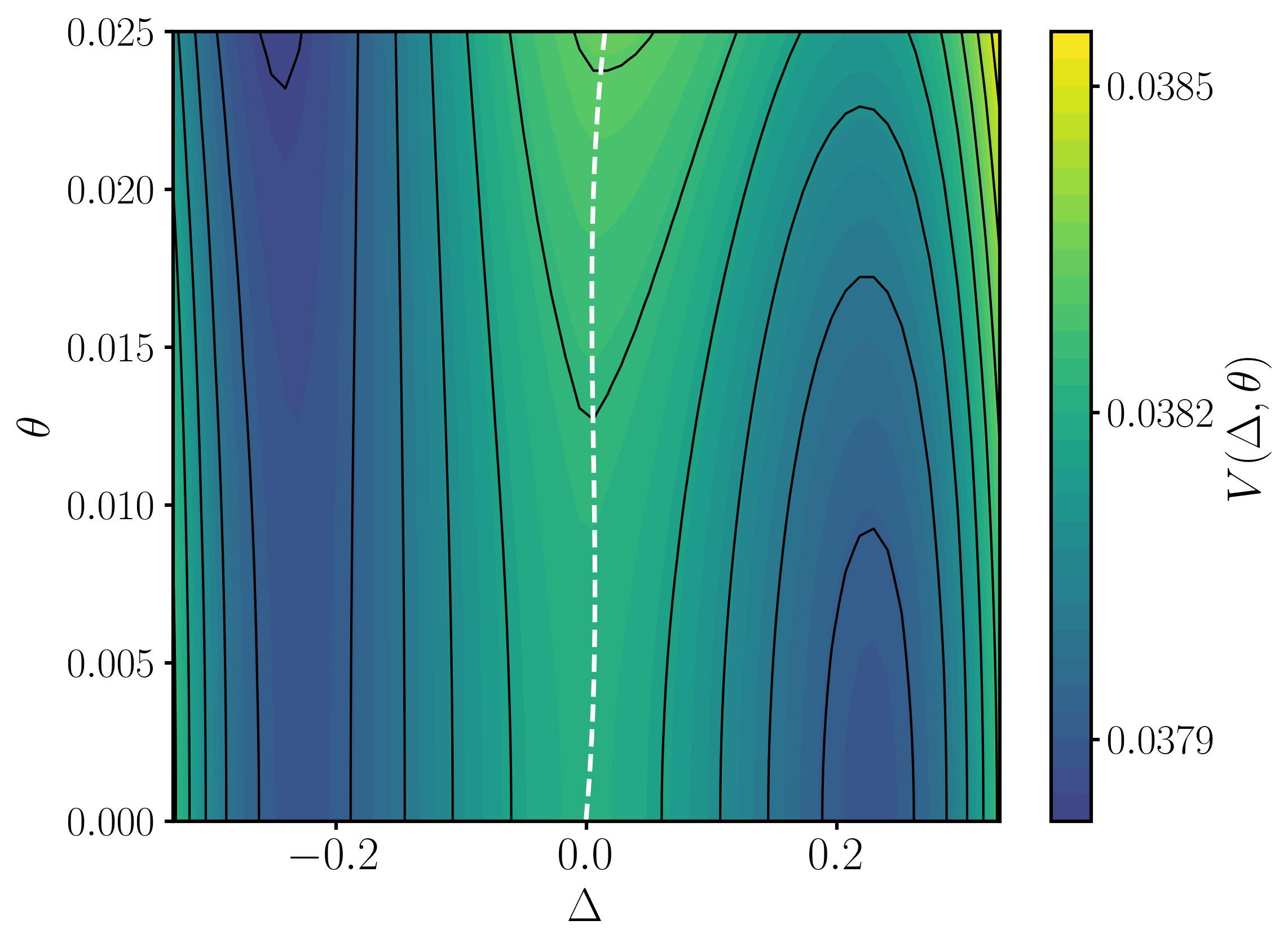}
    \caption{Dependence of Regime 3 on different values of $\theta$, with $c=0$. The white dashed line indicates the location of the intermediate maximum for each value of $\theta$.}
    \label{fig:potential_contour2}
\end{figure}

The \textit{third regime} is observed for $\beta g > 1$ and $\tilde{\Delta}_1 < \Delta_0 < \tilde{\Delta}_2$, as shown in Fig. \ref{fig:potential_regime_3}. This regime exhibits a double-well potential, allowing for distinct behaviors: positive-energy states correspond to zero-mean limiting cycles, while negative-energy states correspond to non-zero-mean limiting cycles. The two minima indicate two possible spin glass freezing points. For $\theta = 0$, the solution aligns with Sompolinsky’s results \cite{Sompolinsky1988}. By introducing $\theta \neq 0$ again breaks the symmetry and shifts the minima, restricting the range of viable autocorrelation functions. This trend is reflected in Fig. \ref{fig:potential_contour2}, where the left minimum deepens and the right minimum becomes shallower as $\theta$ increases.

\begin{figure}[t]
    \includegraphics[width=\linewidth]{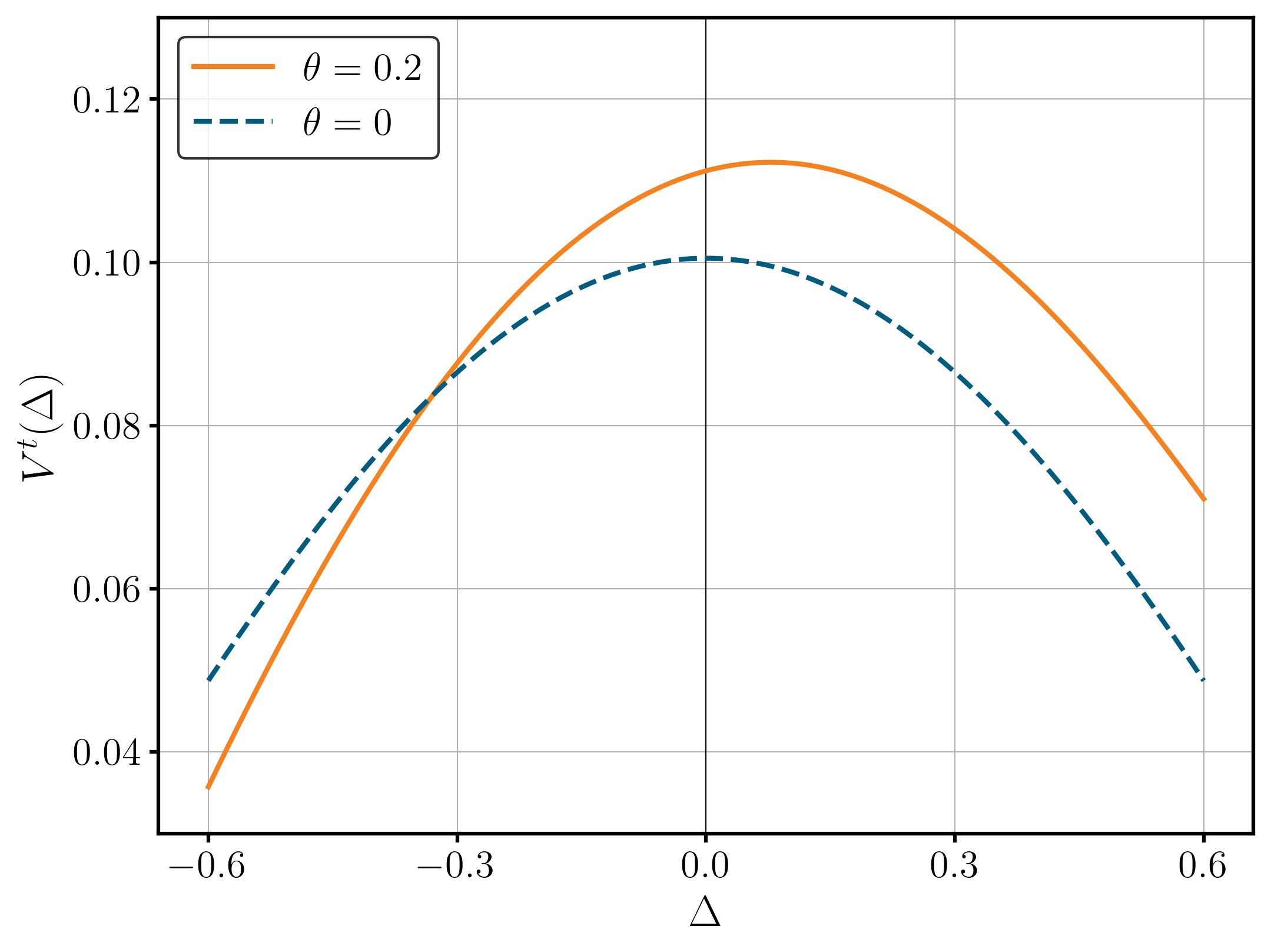}
    \caption{Potentials for Regime 4 for different values of $\theta$.}
    \label{fig:potential_regime_4}
\end{figure}

The \textit{fourth regime} occurs for $\beta g > 1$ and $\Delta_0 > \tilde{\Delta}_2$, as illustrated in Fig. \ref{fig:potential_regime_4}. In this case, the local minima are pushed outside the allowed range of $\Delta$, and the potential qualitatively resembles that of the first regime.

\section{Stability Characteristics}
\label{StabilityAnalysis}

The potential discussed in the above section gives us an idea about the different solutions of interest of our dynamic system. However, a rigorous stability analysis needs to be performed for these points in order to fully grasp the expressivity of our system. As such, the stability is studied via the linear perturbation of the average flow 
\begin{align}
\label{eq:av_flow}
\chi^2(\tau)= 
\lim_{\tilde{\tau}\to\infty}
\frac{1}{N}
\mathbb{E}_N\left[
\sum_{i,j}
\frac{\delta y_i^2(\tau+\tilde{\tau})}{\delta \tilde{y}_j^2(\tau)}
\right],
\end{align}
obtained by adding an infinitesimal source term $\tilde{y}_j$ to Eq. \eqref{eq:y_system}. The stability of the system can then be described by the asymptotic behaviour dictated by Eq. \eqref{eq:av_flow} via its Lyapunov coefficient
\begin{align}
\label{eq:ljupanov}
    \lambda = \lim_{\tau\to\infty} \frac{\ln\left[\chi^2(\tau)\right]}{2\tau}.
\end{align}

Positive Lyapunov exponents result in unstable behaviour, where minor fluctuations, always present due to the stochastic nature of the system, get amplified. Negative Lyapunov exponents on the other hand damp such fluctuations, such that the system remains stable. We can further represent the flow over the modes off the underlying energy-surface through a superposition of elementary states $\chi_n$ resulting in
\begin{align*}
    \chi^2(\tau)= \sum_{n=0}^\infty \chi_n \exp\left(2 \omega_n \tau\right).
\end{align*} 

The frequencies can be expressed  via $\omega_n=-1\pm \sqrt{1-E_n}$, through the energies of a representative effective system. As such, they are obtained via solving the eigenvalue problem of the following Schrödinger-like one-dimensional equation,
\begin{align}
\label{eq:schroedinger}
    \left(-\frac{\partial^2}{\partial \tau^2} - \frac{\partial^2 V}{\partial \Delta^2}\right) \psi_n(\tau) = E_n\psi_n(\tau).
\end{align}

The above formula is a direct result of applying perturbative Martin-Siggia-Rose (MSR) theory to the system in question \cite{tauber2014, Sompolinsky1988, Martin1973}. Further, the Schrödinger-like equation forms an explicit connection between the potential $V$ and the average flow $\chi^2$, thus we are able to calculate \eqref{eq:ljupanov} through the lowest energy eigenstate of Eq. \eqref{eq:schroedinger}. This gives us the Lyapunov exponent through
\begin{align}
\label{eq:ljupanov}
    \lambda = -1 + \sqrt{1-E_0}.
\end{align}

Solving the Schrödinger-like eigenvalue problem for the points of interest discussed in Sec.\ref{sec:auto_regimes} fully characterises our dynamic system.

\subsection{Time-Independent}
For time independent problems we only need to look at the ``fluctuation potential" $W=-\partial^2 V(\Delta)/\partial\Delta^2$ to determine the systems stability. The time-independence is only fulfilled for the extrema of the potential landscapes shown in Sec. \ref{sec:auto_regimes}. The general Formula for the second differentiation of the potential can be obtained through 
\begin{align*}
    \frac{\partial^2 V^t(\Delta)}{\partial \Delta^2}
    &= 
    \beta^2g^2 \lim_{N\to\infty} \mathbf{E}_{N} \Bigg\{
    \int_{-\infty}^\infty \int_{-\infty}^\infty \int_{-\infty}^\infty \mathrm{D}z_3 \, \mathrm{D}z_2 \, \mathrm{D}z_1 \, 
    \\
    &\times
    F'\left[\hat{\alpha}(\tau) z_1 + \hat{\beta}(\tau) z_3 + (\theta_i^t + c_i)\right]
    \\
    &\times
    F'\left[\hat{\alpha}(\tau) z_2 + \hat{\gamma}(\tau) z_3 + (\theta_i^{t} + c_i)\right]
    \Bigg\} - 1,
\end{align*}
with $F'(x) = \beta g\left[1-\tanh^2(\beta g x)\right]$. A detailed derivation to arrive at the formula for the above potential can be found in Appendix~\ref{app;derivationPotentialFluctuation}.

It can be seen from the combinations of Eqs. \eqref{eq:schroedinger} and \eqref{eq:ljupanov}, that, as long as $W=-\frac{\partial^2 V^t(\Delta)}{\partial \Delta^2}>0$ the system is stable. Otherwise, it is unstable for the static points. It is straightforward to see that a lower bound on the fluctuation potential can readily be derived to be $W=1-\beta^4 g^4$. Thus, we see that as long as $\beta g<1$, the single fixed point remains stable. Further, the inclusion of $\theta_i\neq0$ results in a new non-trivial stable spin glass freezing mode.

In general, the fluctuation potential at constant values of $\Delta$ will take the reduced form
\begin{align*}
    W
    &=
    1 - \beta^2g^2 \lim_{N\to\infty} \mathbf{E}_{N} \Bigg\{
    \int_{-\infty}^\infty \mathrm{D}z_3 \, 
    \\
    &
    \nonumber
    \times
    F'\left[\hat{\beta}(\tau) z_3 + (\theta_i^t + c_i)\right]
    F'\left[\hat{\gamma}(\tau) z_3 + (\theta_i^{t} + c_i)\right]
    \Bigg\}.
\end{align*}

This can be further refined by investigating \eqref{eq:pde_autocorrelation} and \eqref{eq:autocorrelation} for the case of $\Delta>0$. Specifically, we see that by re-expressing the above formula through an expression of time independent $\Delta$, we obtain
\begin{align}
    W
    &=
    1 - \beta^4g^4 \lim_{N\to\infty} \mathbf{E}_{N} \Bigg\{
    \int_{-\infty}^\infty \mathrm{D}z_3 \, \Bigg[ 1 +
    \nonumber
    \\
    &
    F^4\left[\beta(\tau) z_3 + (\theta_i^t + c_i)\right]
    -2 \,
    F^2\left[\beta(\tau) z_3 + (\theta_i^t + c_i)\right]   
    \Bigg]
    \Bigg\},
    \nonumber
    \\
    &=
    1-\beta^4 g^4 + \beta^2 g^2 \Delta - \beta^4 g^4
    \nonumber
    \\
    &\times
    \lim_{N\to\infty} \mathbf{E}_{N} \Bigg\{
    \int_{-\infty}^\infty \mathrm{D}z_3 
    F^4\left[\beta(\tau) z_3 + (\theta_i^t + c_i)\right]\Bigg\}
    \label{eq:tid_stability}
\end{align}
where we used that $F'(x) = \beta g\left[1-F^2(x)\right]$. 

Similar results are obtained for the case of $\Delta<0$. We can see from the above equation, that, since the remaining integral cannot fully counterbalance the effect of the $\Delta$ dependent term, that for small values of $\Delta$, we can still obtain $W>0$, i.e. stable regions for the case of $\beta g>1$. These regions however, are only exhibited close to $\beta g=1$. The small values of $0\neq|\Delta|\ll 1$ can be obtained by the central minima under the influence of non-zero $\theta$ and $c$ values. Thus, $\theta$ and $c$ slightly enlarge the stable region and act as a regularization.

We can further derive an upper limit for the stability by approximating the last integral in equation \eqref{eq:tid_stability} through Jensen's inequality of convex functions (here the square of $F(x)$) by
\begin{align*}
    W
    &\leq
    1-\beta^4 g^4 + \beta^2 g^2\Delta -\beta^4 g^4 \Delta^2
    \\
    &=
    1-\beta^2 g^2\left[\beta^2 g^2-\Delta(2-\Delta)\right]  
    \\
    &\leq 1 - \beta^4 g^4 \left[ \Delta - 1 \right].
\end{align*}

Additionally, we used that for the time-independent case we have $\Delta\leq \beta^2 g^2$. This also shows that the stable region, even though it can be slightly extended into the regime of $\beta g>1$, can not exist for $\Delta>1$ anymore.

\subsection{Time-Dependent}
For a time-dependent $\Delta$, both terms of the Schrödinger-like equation \eqref{eq:schroedinger} contribute to the stability analysis. This analysis is not straightforward, however, we can identify one eigenstate of this equation as $\dot{\Delta}(\tau)$. This can be readily verified by differentiating equation \eqref{eq:pde_autocorrelation} with respect to $\tau$, yielding,
\begin{align*}
    \left(-\frac{\partial^2}{\partial \tau^2} - \frac{\partial^2 V}{\partial \Delta^2}\right) \frac{\partial}{\partial \tau} \Delta(\tau)= 0.
\end{align*}

Additionally, we obtain the eigenvalue to be equal to zero. However, it is still unclear whether this eigenstate corresponds to the sought ground state of the system needed to characterize the Lyapunov exponent. It is worth noting that this eigenstate is odd, i.e., $\dot{\Delta}(-\tau)=-\dot{\Delta}(\tau)$, following from the even symmetry of $\Delta(\tau)$.

By further analysing the system, we can conclude that the only potentially stable time-dependent solutions must correspond to either a time-decaying or a time-periodic solution.

For the time-decaying case, the eigenstate $\dot{\Delta}$ has exactly one node at $\tau=0$. This is a direct consequence of $\dot{\Delta}$ being odd. This can be used to identify that this eigenstate needs to be the first excited state of the system, according to the Sturm Oscillation Theorem\cite{Pearson2005}. Thus, the ground state must have an energy below that of $\dot{\Delta}$, resulting in a negative eigenvalue $E_0<0$. Consequently, the time-decaying solutions are unstable.

For the time-periodic solutions, the eigenstates and eigenvalues form a continuous band of solutions. It suffices to note that $\dot{\Delta}$ is an eigenstate with zero energy, so that the next higher excited states in the continuum need to have positive energies. Since the positive and negative eigenvalue states are connected through the continuum, these solutions are also unstable in terms of Lyapunov stability.

\begin{figure}
    \centering
    \includegraphics[width=\linewidth]{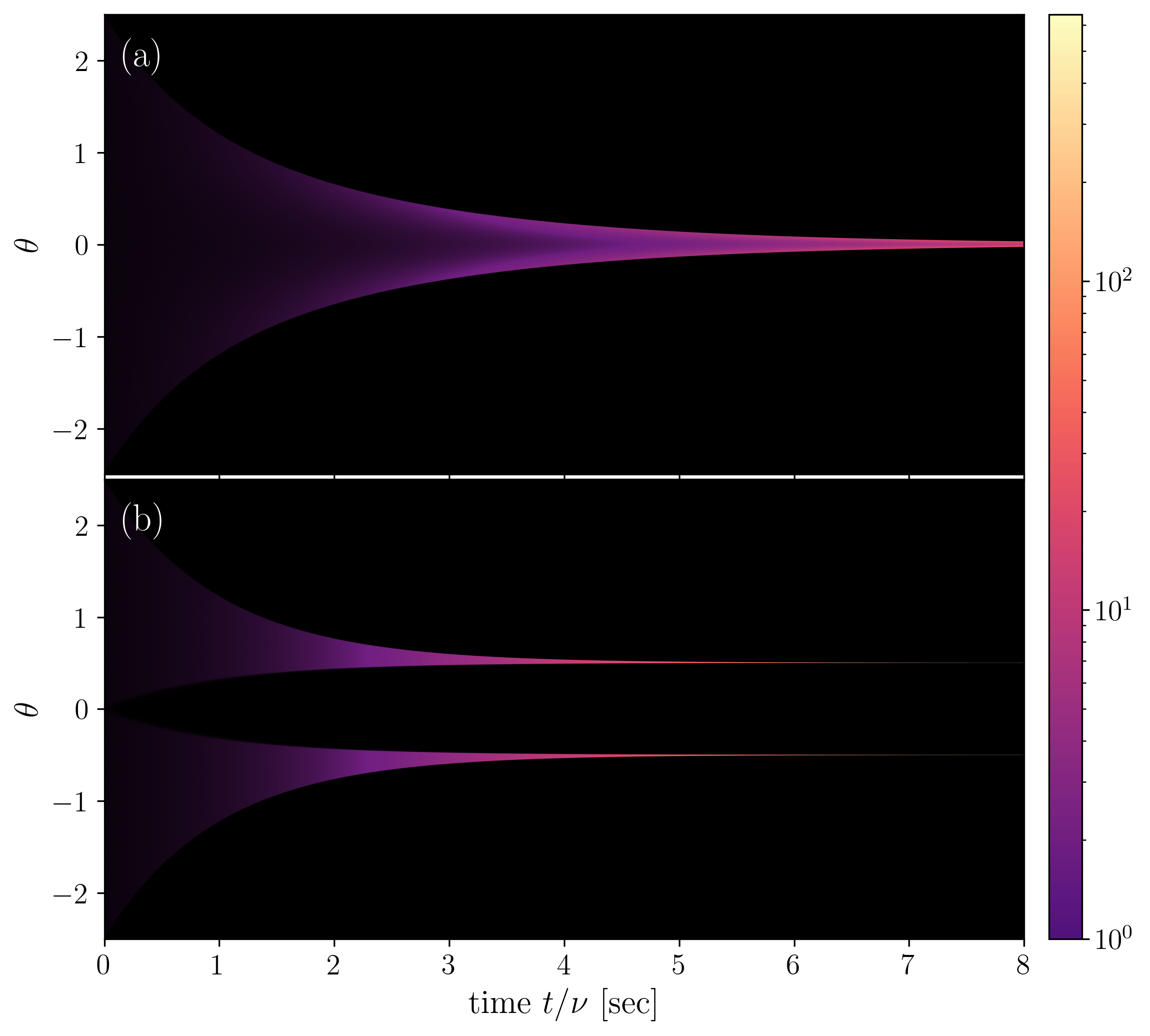}
    \caption{Time evolution of $\theta$ for 1000 agents initially uniformly distributed in the range $\theta\in[-2.5, 2.5]$. The density plots illustrate convergence to either one or two stable values depending on the parameter regime: (a) $\alpha\beta < 1$ and (b) $\alpha\beta > 1$, with $c = 0$.}
    \label{fig:theta_stability}
\end{figure}

\subsection{Analysis of $\theta$ Influence}
The stability analysis so far only implicitly assumed time-independent $\theta$ values. However, the general formulation shown in equation \eqref{eq:ge_theta} is much richer and needs to be taken into account. While the final equation for $\theta_i$ is deterministic, residual coupling to the system variable $y$ may persist, as the separation of timescales is exact only in the limit of infinite timescale separation. Thus, we start with a stability analysis for $\theta_i$:
\begin{align*}
    \frac{\mathrm{d}\theta}{\mathrm{dt}} 
    &=
    \nu \Big\{
    \alpha \tanh\left[\beta(c_i + \theta_i)\right]
    -\theta_i
    \Big\}\\
    &=
    \nu f(\theta_i).
\end{align*}

The first differentiation of the function $f(\theta_i)$ is given by
\begin{align}
    f'(\theta_i)
    &=
    \alpha\beta \Big\{1-\tanh^2\left[\beta \theta_i^0\right] \Big\} - 1 
    \nonumber
    \\
    &\leq \alpha\beta -1.
    \label{eq:theta_stability}
\end{align}

This suggests that in the case of $\alpha\beta<1$, all points that are time-independent are stable. However, since stable and unstable points need to take turns for continuous functions, it follows that in this case we have only one single time-independent (and stable) point. The value of this point depends crucially on $c_i$. In the case of $c_i=0$, it is straightforward to see that $\theta_i=0$ as well.

For the more general case of $\alpha\beta>1$, we obtain the three possible equilibrium points $\theta_-\approx - \alpha<0$ and $\theta_+\approx +\alpha>0$ as well as an intermediate point $\theta_0$ with $\theta^-< \theta^0 < \theta^+$. There are always at least 1 and maximally 3 equilibrium points present. This is dictated by the specific value of $c_i$ as well as the product of $\alpha\beta$. If $c_i>\alpha$, then there is only $\theta^+$, if $c_i<-\alpha$, then there is only $\theta^-$. If $c_i=\pm\alpha$, then there exist the points $\theta^\pm$. If $\alpha<c_i<\alpha$, then all 3 points are present. Thus, we will restrict our analysis to this slightly more constrained case.

Looking at the derivative of the above Eq. \eqref{eq:theta_stability} for the two outer points $\theta^\pm$ we obtain
\begin{align*}
    f'(\theta_i^\pm)
    &=
    \alpha\beta \Big\{1-\tanh^2\left[\beta \theta_i^\pm\right] \Big\} - 1 \\
    &=
    \alpha\beta \Big[1- \left(\frac{\theta_i^\pm}{\alpha}\right)^2 \Big] - 1 \\
    &\approx -1 \\
    &< 0.
\end{align*}

Thus we see that the outer equilibrium points $\theta_i^\pm$ are stable while the inner point is unstable (when $\alpha\beta>1$). This shows that we get a bifurcation for the case $\alpha\beta>1$.
This analysis is further supported by the numerical simulations on the time-evolution of $\theta_i$ shown in Fig. \ref{fig:theta_stability}.

For timescales comparable or bigger than $1/\nu$, the influence of $\theta$ becomes noticeable and our potential exhibits an additional implicit time-dependency. For the stability analysis it makes sense to look at the connection between the original stability for small timescales $\tau$, for which we can assume $\theta_i\equiv const.$, and for extremely long timescales for which $\theta_i$ already reached its saturation point.

In our case, i.e. $\alpha\beta>1$, every initial distribution of $\theta$ converges to either one of the two possible stable fix-points of theta, resulting in the adjusted potential
\begin{align}
\label{eq:pmpotential}
    V^\pm(\theta)
    &=
    -\frac{\Delta^2}{2}+ 
    \lim_{N\to\infty}\mathbf{E}_{N}\Bigg(
    \int_{-\infty}^{\infty}
    \int_{-\infty}^{\infty}
    \int_{-\infty}^{\infty}
    \mathrm{D}z_3
    \mathrm{D}z_2
    \mathrm{D}z_1 \\
    &\times
    \nonumber
    \Bigg\{
    w_+
    \Phi\left[ \sqrt{\Delta_0 - |\Delta|} z_2 + \sqrt{|\Delta|}z_3 + (c_i + \alpha) \right]
    \\
    &\times
    \nonumber
    \Phi\left[ \sqrt{\Delta_0 - |\Delta|} z_1 + \sqrt{|\Delta|}z_3 + \mathrm{sgn}(\Delta) (c_i + \alpha) \right]\\
    &
    \nonumber
    +
    w_-
    \Phi\left[ \sqrt{\Delta_0 - |\Delta|} z_1 + \sqrt{|\Delta|}z_3 + (c_i - \alpha) \right]
    \\
    &\times
    \nonumber
    \Phi\left[ \sqrt{\Delta_0 - |\Delta|} z_2 + \sqrt{|\Delta|}z_3 + \mathrm{sgn}(\Delta) (c_i - \alpha) \right]
    \Bigg\}
    \Bigg),
\end{align}
for all $\Delta$ and $w_++w_- \equiv 1$. A more detailed derivation can be found in Appendix~\ref{app:proofPotentialPm}.

The only significant difference in comparison to the potential used for the derivation of the constant $\theta$ stability analysis is the magnitude of the value of $\theta$. Therefore, since the only real magnitude depending derivation corresponds to the time-independent fix-point of $\Delta$ close to the origin, we can conclude that only in this case, the stable region will become unstable. This is a direct consequence of $\theta$ controlling the value of $\Delta$ near the origin. 

It is worth noting that, if $\alpha\beta>>1$, then only one single extrema for the $\theta_i$ values will remain, simplifying the above potential even further by setting one if the frequencies to zero. This is similar to the case where $\alpha\beta<1$. However, in this case the equilibrium lies somewhere between $-\alpha$ and $\alpha$ as long as $c_i\ne 0$. 
In the case of $\alpha\beta<1$ and $c_i=0$ for all agents, we regain the same potential as Sompolinsky derived in 1988 \cite{Sompolinsky1988}.

\section{Summary and Discussion}\label{Summary}
In this work, we have explored the construction of a potential landscape in GRNs with the inclusion of an epigenetic feedback term, which is essential for modelling cellular reprogramming processes~\cite{KanekoPRR2020,kaneko2007evolution}. The motivation behind this study stems from the increasing significance of computational biology, particularly in the analysis of scRNA-seq data~\cite{Klein1}, and the integration of machine learning techniques with mathematical models to predict cellular differentiation~\cite{Briscoe1,Briscoe2,Raju1,Raju2,Raju3}. To this end, the Waddington metaphor has emerged as a promising direction for constructing a prototype mathematical framework, particularly through bifurcation analysis and the study of the nature of fixed points, which often extend to real-world data~\cite{Waddington,Briscoe1,Briscoe2,Raju1,Raju2,Raju3}. The Waddington metaphor focuses on the evolution of cells on a complex potential surface, which is typically shaped by GRNs~\cite{Waddington,Briscoe1,Briscoe2,Raju1,Raju2,Raju3}.

In our approach, we have merged the classical problem of constructing a potential well with the Waddington metaphor, extending it to a slightly deformed version when various perturbations--such as feedback mechanisms--are present in the network~\cite{KanekoPRR2020,kaneko2007evolution,matsushita2022dynamical,miyamoto2015pluripotency}. These perturbations are key to analysing the stability of fixed points and the nature of bifurcations, which are critical for understanding cellular differentiation processes and disease progression, such as cancer metastasis~\cite{Mohit,Mohit2}. Our study begins with this motivation, focusing on the construction of a potential landscape when an epigenetic feedback term is included in the GRN. This feedback term is typically used to describe the reprogramming process in GRNs, which is crucial for cellular reprogramming or repair~\cite{KanekoPRR2020,kaneko2007evolution,matsushita2022dynamical,miyamoto2015pluripotency}.

The concept is grounded in the well-known Hopfield model, a popular prototype of GRNs, where the construction of such a potential landscape was first provided by Sompolinsky more than three decades ago using DMFT. We replicated similar analysis for a slightly deformed version of the Hopfield model, proposed by Kaneko, and demonstrated that the potential landscape can still be constructed through a transformation back into the non-deformed Hopfield model.

The complexity arises when perturbations or feedback terms are introduced into the system, adding extra dimensions to the biophysical model. This makes the construction of the potential landscape more challenging, as the additional dimensions complicate the mathematical analysis. In this work, we develop a mathematical mechanism to handle these feedback terms, which describe reprogramming in GRNs, to construct the potential landscape. This development allows us to analyse the stability of global dynamics.

We also compare the deformation of the potential landscape caused by the feedback term with the non-feedback case using the DMFT framework. 
The focus of our DMFT analysis is to observe how the autocorrelation function and the matching moments are affected by varying the strength of the feedback term. \emph{Our results reaveal that even small changes in the feedback term can lead to significant deformations in the potential landscape.}

While the method has certain limitations--given the mathematical complexity and the length of the analysis--it provides a valuable framework for constructing a reliable version of the potential landscape of GRNs with feedback. This approach will allow us to analyse the strength of feedback and its effect on the potential surface, providing insight into various intrinsic parameters.

As a future direction, it would be intriguing to examine the nature of the potential landscape in the presence of implicit and explicit time-dependent feedbacks~\cite{matsushita2023generic,matsushita2022dynamical}. Understanding the regulation of GRNs through potential surfaces and the stability properties of the corresponding dynamical systems will help bridge the gap between theoretical models and experimental data. This could allow for the direct analysis of stability properties and the construction of accurate potential surfaces for various real biological datasets.


\section*{Acknowledgements}
We acknowledge that challenging academic experiences can sometimes lead to unexpected collaborations and new perspectives. In this spirit, SH and SS recognize the broader research journey--including an earlier period associated with Prof. Heinz Koeppl (TU Darmstadt)--that ultimately enabled the completion of this work. SS acknowledges Archishman Raju for insightful discussions and critical suggestions, without which the project would have been incomplete. SS also acknowledges Yuuki Matsushita and Kunihiko Kaneko for valuable discussions. We acknowledge ChatGPT for grammatical corrections and improvements.


\section*{Data Availability}
The data that support the findings of this study are available
within the article.

\appendix
\onecolumngrid
\section{Treating Noise Expectation}
\label{app:treating_noise_expectation}

In this section, we present a detailed derivation of the noise expectation defined in Eq.~\eqref{eq:noise_expectation}. The objective is to evaluate the expectation value of the noise-dependent functional $y_1(t, \eta(t))$. First, we introduce the shorthand
\[
\alpha = \frac{2 i \beta \log(u)}{\pi},
\]
to simplify the derivation. Now, we can express $H(u,t)$ as a simple power series:
\begin{align*}
    H(u,t)
    &=
    \sum_{n=0}^\infty
    \frac{\alpha^n}{n!} 
    \mathbb{E}_{\eta} \left\{
    y_1^n(t, \eta(t))
    \right\} \\
    &=
    \sum_{n=0}^\infty
    \frac{\alpha^n}{n!} 
    \mathbb{E}_{\eta} \left\{
    \left[\int_0^t \eta(\tau)\exp\left(\tau-t\right) \mathrm{d}\tau\right]^n
    \right\} \\
    &=
    \sum_{n=0}^\infty
    \frac{\alpha^n}{n!} 
    \int_0^t
    ...
    \int_0^t
    \mathbb{E}_{\eta} \left\{
    \prod_{j=1}^{n}
    \eta(\tau_j)
    \right\}
    \exp\left(\sum_{j=1}^n\tau_j-nt\right) \mathrm{d}\tau_1
    ...
    \mathrm{d}\tau_n .
\end{align*}

At this point, we make use of the assumption that the underlying noise process is Gaussian. As a consequence, all odd-order moments vanish, and only even orders contribute to the expansion. This reduces the expression to
\begin{align*}
    H(u, t)
    &=
    \sum_{n=0}^\infty
    \frac{\alpha^{2n}}{(2n)!} 
    \int_0^t
    ...
    \int_0^t
    \mathbb{E}_{\eta} \left\{
    \prod_{j=1}^{2n}
    \eta(\tau_j)
    \right\}
    \exp\left( \sum_{j=1}^{2n}\tau_j-2nt\right) \mathrm{d}\tau_2
    ...
    \mathrm{d}\tau_{2n} \\
    & =
    \label{eq:isserlis_theorem_asssumption}
    \sum_{n=0}^\infty
    \frac{\alpha^{2n}}{(2n)!} 
    \int_0^t
    ...
    \int_0^t
    \left\{
    \sum_{p\in\text{permut.}}
    \prod_{i,j\in p}
    g^2 C(\tau_i-\tau_j)
    \right\}
    \exp\left( \sum_{j=1}^{2n}\tau_j-2nt\right) \mathrm{d}\tau_2
    ...
    \mathrm{d}\tau_{2n} \\
    & =
    \sum_{n=0}^\infty
    \frac{\left( g \alpha\right)^{2n}}{(2n)!} 
    \sum_{p\in\text{permut.}}
    \int_0^t
    ...
    \int_0^t
    \left\{
    \prod_{i,j\in p}
    C(\tau_i-\tau_j)
    \exp\left(
    \tau_i + \tau_j - 2t\right)
    \right\}
    \mathrm{d}\tau_2
    ...
    \mathrm{d}\tau_{2n}.
\end{align*}

In the second step above, Isserlis’ theorem~\cite{isserlis1918} was employed to express higher-order moments of the Gaussian noise in terms of pairwise correlations involving the autocorrelation function $C(\tau_i-\tau_j)$.  

At this stage, the explicit labelling of the integration variables becomes immaterial: the indices no longer carry any additional information, as each term in the product is identical up to permutation. We may therefore treat them as dummy indices and drop their explicit distinction. This leads to the reduced expression
\begin{align*}
    H(u, t)
    & =
    \sum_{n=0}^\infty
    \frac{\left(g \alpha\right)^{2n}}{(2n)!} 
    \sum_{p\in\text{permut.}}
    \prod_{i,j\in p}
    \left\{
    \int_0^t
    \int_0^t
    C(\tau_i-\tau_j)
    \exp\left(
    \tau_i + \tau_j - 2t\right)
    \mathrm{d}\tau_i
    \mathrm{d}\tau_j
    \right\} \\
    & =
    \sum_{n=0}^\infty
    \frac{\left(g \alpha\right)^{2n}}{(2n)!} 
    \sum_{p\in\text{permut.}}
    \prod_{i,j\in p}
    \left\{
    2
    \int_0^t
    \int_{-\tau}^{\tau}
    C_{\eta}(s)
    \exp\left(
    \tau - 2t\right)
    \mathrm{d}s
    \mathrm{d}\tau
    \right\} \\
     & =
    \sum_{n=0}^\infty
    \frac{\left(g \alpha\right)^{2n}}{n!} 
    \underbrace{
    \left\{
    \int_0^t
    \exp\left(
    \tau - 2t\right)
    \left[
    \int_0^{\tau}
    C_{\eta}(s)
    \mathrm{d}s
    \right]
    \mathrm{d}\tau
    \right\}^n
    }_{
    \tilde{C}^n(t)
    } \\
    & =
    \exp\left[\alpha^2 g^2 \tilde{C}(t)\right].
\end{align*}

In the final steps, symmetry arguments and a change of integration variables were used to decouple the double integrals and isolate the contribution of the noise autocorrelation function. The resulting expression shows that the noise expectation can be written in closed form as an exponential involving the effective correlation measure $\tilde{C}(t)$.

\section{Noise Self-Consistency}
\label{app:self_consistency}

In this section, we derive the self-consistency relations for the effective noise process that appears in the dynamical mean-field description. These relations connect the statistics of the effective noise to the autocorrelation function of the transformed network activity and justify the expressions used in the main text.

We begin by considering the mean of the effective noise. By definition, the noise term can be written as a weighted sum over the network interactions, such that,
\begin{align*}
   \mathbf{E}_{t,\eta} \left[\eta_i(t)\right]
   =
   \lim_{N\to\infty}
   \mathbf{E}_{t,J} \left\{
   \sum_{j=1}^N J_{ij} F[y_j(t)]
   \right\}.
\end{align*}

Since the couplings are drawn independently with zero mean, $\mathbf{E}_{J} \left[J_{ij}\right]=0$, the expectation value vanishes as well, $\mathbf{E}_{t,\eta} \left[\eta_i(t)\right] = 0$. We next determine the temporal correlations of the effective noise. Starting from its definition, we obtain,
\begin{align*}
    \mathbf{E}_{t,\eta} \left[\eta_i(t) \eta_j(t+\tau)\right]
    &=
    \lim_{N\to\infty}
    \mathbf{E}_{t,J} \left\{
    \sum_{l=1}^N J_{il}
    \sum_{k=1}^N J_{jk}
    F[y_l(t)] F[y_k(t+\tau)]
    \right\}.
\end{align*}

Using the identity $\mathbf{E}_{J} \left[J_{il}J_{jk}\right]=g^2 \delta_{ij}\delta_{lk}/N$ and the independence of the couplings, the disorder average can be evaluated explicitly, yielding
\begin{align*}
    \mathbf{E}_{t,\eta} \left[\eta_i(t) \eta_j(t+\tau)\right]
    &=
    \lim_{N\to\infty} \sum_{l=1}^N \sum_{k=1}^N \mathbf{E}_{J} \left[  J_{il} J_{jk} \right] \mathbf{E}_{t} \left\{ F[y_l(t)] F[y_k(t+\tau)]\right\}
    \\
    &=
    \lim_{N\to\infty} 
    \delta_{ij} g^2 \frac{1}{N}\sum_{k=1}^N \mathbf{E}_{t} \left\{ F[y_k(t)] F[y_k(t+\tau)]\right\}\\
    &=
    \delta_{ij} g^2 C(\tau),
\end{align*}
with $ C(\tau) \equiv C_\infty(\tau) $. This establishes the self-consistent relation between the effective noise autocorrelation and the population-averaged autocorrelation function used in the main text.

\section{Proof of Potential $V^t(\Delta)$}
\label{app:proof_potential}
 In what follows, we will derive the the specific equations concerning different forms of the potential $V^t(\Delta)$.

\subsection{Proof that $V^t(\Delta)$ acts as a potential for $\Delta$}
\label{app:proof_potential}

In this appendix, we provide a detailed derivation showing that the effective potential $V^t(\Delta)$, defined in Eq.~\eqref{eq:effective_potential}, indeed satisfies Eq.~\eqref{eq:pde_autocorrelation_potential}.  
Our goal is to explicitly compute the derivative of $V^t(\Delta)$ with respect to $\Delta$ and show that it reproduces the right-hand side of Eq.~\eqref{eq:pde_autocorrelation}.

Starting from the definition, we have
\begin{align}
    \frac{\partial V^t(\Delta)}{\partial \Delta}
    &=
    -\Delta+ 
    \frac{\partial}{\partial \Delta}
    \lim_{N\to\infty} \mathbf{E}_{N} \Bigg\{
    \int_{-\infty}^{\infty}
    \int_{-\infty}^{\infty}
    \int_{-\infty}^{\infty}
    \mathrm{D}z_3
    \mathrm{D}z_2
    \mathrm{D}z_1
    \Phi\Big[ \sqrt{\Delta_0 - |\Delta|} z_2 + \sqrt{|\Delta|}z_3 + (\theta_i^t + c_i)\Big]
    \nonumber
    \\
    &\times
    \Phi\Big[ \sqrt{\Delta_0 - |\Delta|} z_1 + \sqrt{|\Delta|}z_3 + \mathrm{sgn}(\Delta) (\theta_i^t + c_i)\Big]\Bigg\}
    \nonumber
    \\
    &=
    -\Delta+ 
    \lim_{N\to\infty} \mathbf{E}_{N} \Bigg(
    \int_{-\infty}^{\infty}
    \int_{-\infty}^{\infty}
    \int_{-\infty}^{\infty}
    \mathrm{D}z_3
    \mathrm{D}z_2
    \mathrm{D}z_1
    \mathrm{sgn}(\Delta)
    \Bigg[
    -\frac{1}{2 \sqrt{|\Delta|}} z_3
    +
    \frac{1}{2 \sqrt{\Delta_0-|\Delta|}} z_2
    \Bigg]
    \\
    &\times
    \Bigg\{
    \Phi\Big[ \sqrt{\Delta_0 - |\Delta|} z_1 + \sqrt{|\Delta|}z_3 + (\theta_i^t + c_i)\Big]
    F\Big[ \sqrt{\Delta_0 - |\Delta|} z_2 + \sqrt{|\Delta|}z_3 + \mathrm{sgn}(\Delta) (\theta_i^t + c_i)\Big]
    \nonumber
    \\
    &+
    F\Big[ \sqrt{\Delta_0 - |\Delta|} z_2 + \sqrt{|\Delta|}z_3 + (\theta_i^t + c_i)\Big]
    \Phi\Big[ \sqrt{\Delta_0 - |\Delta|} z_1 + \sqrt{|\Delta|}z_3 + \mathrm{sgn}(\Delta) (\theta_i^t + c_i)\Big]
    \Bigg\}
    \Bigg)
    \label{eq:Vt_indicator}
\end{align}
where we used that $F(x) = \mathrm{d}\Phi(x)/\mathrm{d}x$.  

To simplify the computation, we split the integral into two parts, corresponding to the two terms inside the square brackets in Eq.~\eqref{eq:Vt_indicator}.  
For the first term, we obtain
\begin{align}
\label{eq:lemmastein1}
    \tilde{I}_1
    &= 
    \int_{-\infty}^{\infty}\, \mathrm{D}z_1 F\left[ \sqrt{\Delta_0 - |\Delta|} z_1 + \sqrt{|\Delta|}z_3 + (\theta_i^t +c_i)\right] \frac{1}{\sqrt{\Delta_0 - |\Delta|}} z_1
    \nonumber
    \\
    &=
    A^{-1} \int_{-\infty}^{\infty}\, \mathrm{D}x F\left[ A x + B\right] x
    \nonumber
    \\
    &=
    A^{-1} \mathbb{E}\left\{ F\left[ A x + B\right] x \right\}
    \nonumber
    \\
    &=
    A^{-1} g^2\mathbb{E}\left\{A F'\left[ A x + B\right] \right\} 
    \nonumber
    \\
    &=
    g^2 \mathrm{sgn}(\Delta) \int_{-\infty}^{\infty} \mathrm{D}z_1\, F'\left[ \sqrt{\Delta_0 - |\Delta|} z_1 + \sqrt{|\Delta|}z_3 + (\theta_i^t +c_i)\right].
\end{align}
The central step here is the use of \textit{Stein's Lemma} \cite{ChenGoldsteinShao2011}. We note that the sign factor $\mathrm{sgn}(\Delta)$ does not influence the derivation even if it is not included in the function argument.

Similarly, for the second term, we have
\begin{align}
\label{eq:lemmastein2}
    \tilde{I}_2
    &=
    \int_{-\infty}^{\infty} \mathrm{D}z_3 \, \Phi\left[ \sqrt{\Delta_0 - |\Delta|} z_2 + \sqrt{|\Delta|}z_3 + (\theta_i^t + c_i)\right] F\left[ \sqrt{\Delta_0 - |\Delta|} z_1 + \sqrt{|\Delta|}z_3 + \mathrm{sgn}(\Delta) (\theta_i^t + c_i)\right] \frac{1}{ \sqrt{|\Delta|}} z_3 
    \nonumber
    \\
    &=
    B^{-1} \int_{-\infty}^{\infty} \mathrm{D}z_3 \, \Phi\left[ A_2 + B z_3 + C\right] F\left[ A_1 + B z_3 + \mathrm{sgn}(\Delta) C\right] z_3
    \nonumber
    \\
    &=
    B^{-1} \mathbb{E}\left\{ \Phi\left[ A_2 + B x + C\right] F\left[ A_1 + B x + \mathrm{sgn}(\Delta) C\right] x \right\}
    \nonumber
    \\
    &=
    g^2 B^{-1} \mathbb{E}\left\{ B F\left[ A_2 + B x + C\right] F\left[ A_1 + B x + \mathrm{sgn}(\Delta) C\right] + B \Phi\left[ A_2 + B x + C\right] F'\left[ A_1 + B x + \mathrm{sgn}(\Delta) C\right] \right\}
    \nonumber
    \\
    &=
    g^2 \int_{-\infty}^{\infty} \mathrm{D}z_3 \, \Bigg\{ F\left[ \sqrt{\Delta_0 - |\Delta|} z_2 + \sqrt{|\Delta|}z_3 + (\theta_i^t + c_i)\right] F\left[ \sqrt{\Delta_0 - |\Delta|} z_1 + \sqrt{|\Delta|}z_3 + \mathrm{sgn}(\Delta) (\theta_i^t + c_i)\right] 
    \nonumber
    \\
    & +
    \Phi\left[ \sqrt{\Delta_0 - |\Delta|} z_2 + \sqrt{|\Delta|}z_3 + (\theta_i^t + c_i)\right] F'\left[ \sqrt{\Delta_0 - |\Delta|} z_1 + \sqrt{|\Delta|}z_3 + (\theta_i^t + c_i)\right] \Bigg\}.
\end{align}
Analogous results are obtained when the arguments of $\Phi$ and $F$ are exchanged.  
Combining these results, we can rewrite the derivative of the potential as
\begin{align*}
    \frac{\partial V^t(\Delta)}{\partial \Delta}
    &=
    -\Delta+ g^2\frac{\mathrm{sgn}(\Delta)}{2} \lim_{N\to\infty} \mathbf{E}_{N} \Bigg\{ \int_{-\infty}^{\infty} \int_{-\infty}^{\infty} \int_{-\infty}^{\infty} \mathrm{D}z_3 \mathrm{D}z_2 \mathrm{D}z_1\\
    &\times
    \Bigg\{ \Phi\left[ \sqrt{\Delta_0 - |\Delta|} z_1 + \sqrt{|\Delta|}z_3 + (\theta_i^t + c_i)\right] F'\left[ \sqrt{\Delta_0 - |\Delta|} z_2 + \sqrt{|\Delta|}z_3 + \mathrm{sgn}(\Delta) (\theta_i^t + c_i)\right] \\
    & +
    F'\left[ \sqrt{\Delta_0 - |\Delta|} z_2 + \sqrt{|\Delta|}z_3 + (\theta_i^t + c_i)\right] \Phi\left[ \sqrt{\Delta_0 - |\Delta|} z_1 + \sqrt{|\Delta|}z_3 - (\theta_i^t + c_i)\right] \Bigg\} \\
    &+
    \Bigg\{ F\left[ \sqrt{\Delta_0 - |\Delta|} z_1 + \sqrt{|\Delta|}z_3 + (\theta_i^t + c_i)\right] F\left[ \sqrt{\Delta_0 - |\Delta|} z_2 + \sqrt{|\Delta|}z_3 + \mathrm{sgn}(\Delta)(\theta_i^t + c_i)\right] \\
    &+
    F\left[ \sqrt{\Delta_0 - |\Delta|} z_2 + \sqrt{|\Delta|}z_3 + (\theta_i^t + c_i)\right] F\left[ \sqrt{\Delta_0 - |\Delta|} z_1 + \sqrt{|\Delta|}z_3 + \mathrm{sgn}(\Delta)(\theta_i^t + c_i)\right] \Bigg\} \\
    &-
    \Bigg\{ \Phi\left[ \sqrt{\Delta_0 - |\Delta|} z_1 + \sqrt{|\Delta|}z_3 + (\theta_i^t + c_i)\right] F'\left[ \sqrt{\Delta_0 - |\Delta|} z_2 + \sqrt{|\Delta|}z_3 + \mathrm{sgn}(\Delta)(\theta_i^t + c_i)\right] \\
    &+
    F'\left[ \sqrt{\Delta_0 - |\Delta|} z_2 + \sqrt{|\Delta|}z_3 + (\theta_i^t + c_i)\right] \Phi\left[ \sqrt{\Delta_0 - |\Delta|} z_1 + \sqrt{|\Delta|}z_3 + \mathrm{sgn}(\Delta)(\theta_i^t + c_i)\right] \Bigg\} \Bigg\} \\
    &=
    -\Delta+ g^2 \mathrm{sgn}(\Delta) \lim_{N\to\infty} \mathbf{E}_{N} \Bigg\{ \int_{-\infty}^{\infty} \int_{-\infty}^{\infty} \int_{-\infty}^{\infty} \mathrm{D}z_3 \mathrm{D}z_2 \mathrm{D}z_1 \\
    &
    F\left[ \sqrt{\Delta_0 - |\Delta|} z_1 + \sqrt{|\Delta|}z_3 + (\theta_i^t + c_i)\right] F\left[ \sqrt{\Delta_0 - |\Delta|} z_2 + \sqrt{|\Delta|}z_3 + \mathrm{sgn}(\Delta)(\theta_i^t + c_i)\right] \Bigg\} \\
    &=
    -\Delta+ g^2\lim_{N\to\infty} \mathbf{E}_{N} \Bigg\{ \int_{-\infty}^{\infty} \int_{-\infty}^{\infty} \int_{-\infty}^{\infty} \mathrm{D}z_3 \mathrm{D}z_2 \mathrm{D}z_1 \\
    &
    F\left[ \sqrt{\Delta_0 - |\Delta|} z_1 + \sqrt{|\Delta|}z_3 + (\theta_i^t + c_i)\right] F\left[ \sqrt{\Delta_0 - |\Delta|} z_2 + \mathrm{sgn}(\Delta) \sqrt{|\Delta|}z_3 +(\theta_i^t + c_i)\right] \Bigg\}
\end{align*}
This establishes that the derivative of $V^t(\Delta)$ coincides with the right-hand side of Eq.~\eqref{eq:pde_autocorrelation}, completing the proof that $V^t(\Delta)$ indeed acts as a potential for $\Delta$.

\subsection{Derivation of the fluctuation potential $W$}
\label{app;derivationPotentialFluctuation}

The fluctuation potential is defined as
\[
W = -\frac{\partial^2 V^t(\Delta)}{\partial \Delta^2}.
\]

We start by writing out the explicit form of the second derivative of the potential:
\begin{align*}
    \frac{\partial^2 V^t(\Delta)}{\partial \Delta^2}
    &=
    -1 - \frac{\beta^2g^2}{2}
    \lim_{N\to\infty} \mathbf{E}_{N} \Bigg( \int_{-\infty}^\infty \int_{-\infty}^\infty \int_{-\infty}^\infty \mathrm{D}z_3 \, \mathrm{D}z_2 \, \mathrm{D}z_1 \, 
    \\
    &\times
    \Bigg\{
    F'\left[\hat{\alpha}(\tau) z_1 + \hat{\beta}(\tau) z_3 + \left(\theta_i^t + c_i\right)\right]
    F\left[\hat{\alpha}(\tau) z_2 + \hat{\gamma}(\tau) z_3 + \left(\theta_i^{t} + c_i\right)\right]
    \left[\frac{\mathrm{sgn}(\Delta)}{\hat{\alpha}(\tau)} z_1 - \frac{1}{\hat{\beta}(\tau)} z_3\right]
    \\
    &
    + \mathrm{sgn}(\Delta)
    F\left[\hat{\alpha}(\tau) z_1 + \hat{\beta}(\tau) z_3 + \left(\theta_i^t + c_i\right)\right]
    F'\left[\hat{\alpha}(\tau) z_2 + \hat{\gamma}(\tau) z_3 + \left(\theta_i^{t} + c_i\right)\right]
    \left[\frac{1}{\hat{\alpha}(\tau)} z_2 - \frac{1}{\hat{\beta}(\tau)} z_3\right]
    \Bigg\}
    \Bigg)\\
        &=
    -1 - \frac{\beta^2g^2}{2}
    \lim_{N\to\infty} \mathbf{E}_{N} \Bigg\{\int_{-\infty}^\infty \int_{-\infty}^\infty \int_{-\infty}^\infty \mathrm{D}z_3 \, \mathrm{D}z_2 \, \mathrm{D}z_1 \, 
    \\
    &\times
    \Big\{
    F'\left\{\hat{\alpha}(\tau) z_1 + \hat{\beta}(\tau) z_3 + \left[\theta_i^t + c_i\right]\right\}
    F\left\{\hat{\alpha}(\tau) z_2 + \hat{\gamma}(\tau) z_3 + \left[\theta_i^{t} + c_i\right]\right\}
    \frac{\mathrm{sgn}(\Delta)}{\hat{\alpha}(\tau)} z_1
    \\
    &
    -
    F'\left\{\hat{\alpha}(\tau) z_1 + \hat{\beta}(\tau) z_3 + \left[\theta_i^t + c_i\right]\right\}
    F\left\{\hat{\alpha}(\tau) z_2 + \hat{\gamma}(\tau) z_3 + \left[\theta_i^{t} + c_i\right]\right\}
    \frac{1}{\hat{\beta}(\tau)} z_3
    \\
    &
    + \mathrm{sgn}(\Delta)
    F\left\{\hat{\alpha}(\tau) z_2 + \hat{\beta}(\tau) z_3 + \left[\theta_i^t + c_i\right]\right\}
    F'\left\{\hat{\alpha}(\tau) z_1 + \hat{\gamma}(\tau) z_3 + \left[\theta_i^{t} + c_i\right]\right\}
    \frac{1}{\hat{\alpha}(\tau)} z_1 
    \\
    &
    -
    \mathrm{sgn}(\Delta)
    F\left\{\hat{\alpha}(\tau) z_2 + \hat{\beta}(\tau) z_3 + \left[\theta_i^t + c_i\right]\right\}
    F'\left\{\hat{\alpha}(\tau) z_1 + \hat{\gamma}(\tau) z_3 + \left[\theta_i^{t} + c_i\right]\right\}
    \frac{1}{\hat{\beta}(\tau)} z_3
    \Big\}
    \Bigg\}
\end{align*}
The evaluation of these integrals can be performed in analogy to Eqs.~\eqref{eq:lemmastein1} and \eqref{eq:lemmastein2}. By applying \textit{Stein's Lemma} \cite{ChenGoldsteinShao2011}, the above expression simplifies to the final form:
\begin{align*}
    \frac{\partial^2 V^t(\Delta)}{\partial \Delta^2}
    &=
    -1 + \beta^2g^2
    \lim_{N\to\infty} \mathbf{E}_{N} \Bigg\{\int_{-\infty}^\infty \int_{-\infty}^\infty \int_{-\infty}^\infty \mathrm{D}z_3 \, \mathrm{D}z_2 \, \mathrm{D}z_1 \, 
    \\
    &\times
    F'\left\{\hat{\alpha}(\tau) z_1 + \hat{\beta}(\tau) z_3 + \left[\theta_i^t + c_i\right]\right\}
    F'\left\{\hat{\alpha}(\tau) z_2 + \hat{\gamma}(\tau) z_3 + \left[\theta_i^{t} + c_i\right]\right\}
    \Bigg\}.
\end{align*}
This completes the derivation of the fluctuation potential $W$.  
The above expression can now be used to analyse the stability of the system at constant $\Delta$.

\subsection{Derivation of $V^\pm(\theta)$}
\label{app:proofPotentialPm}

In this appendix, we derive the potential on timescales that are much longer than $1/\nu$, i.e., timescales sufficiently long for $\theta_i$ to reach its extreme values. Accordingly, we are interested in the long-time limit of the potential,
\begin{align*}
V^\pm(\Delta)\equiv\lim_{t\to\infty}V^t(\Delta).
\end{align*}

To simplify the analysis, we first note that the final values of $\theta_i$ can attain at most two distinct values, which we denote by $\theta^\pm$. With this observation, the potential can be expressed as
\begin{align*}
    V^\pm(\theta)
    &=
    -\frac{\Delta^2}{2}+ 
    \lim_{N\to\infty} \mathbf{E}_{N} \Bigg\{
    \int_{-\infty}^{\infty}\int_{-\infty}^{\infty}\int_{-\infty}^{\infty}
    \mathrm{D}z_3 \mathrm{D}z_2 \mathrm{D}z_1
    \Phi\left[ \sqrt{\Delta_0 - |\Delta|} z_2 + \sqrt{|\Delta|}z_3 + \theta_i^\pm \right]
    \\
    &\times
    \Phi\left[ \sqrt{\Delta_0 - |\Delta|} z_1 + \sqrt{|\Delta|}z_3 + \mathrm{sgn}(\Delta) (\theta_i^\pm +c_i)\right]\Bigg\} \\
    &=
    -\frac{\Delta^2}{2}+ 
    \lim_{N\to\infty} \mathbf{E}_{N} \Bigg(
    \int_{-\infty}^{\infty}\int_{-\infty}^{\infty}\int_{-\infty}^{\infty}
    \mathrm{D}z_3 \mathrm{D}z_2 \mathrm{D}z_1
    \Bigg\{
    w_+
    \Phi\left[ \sqrt{\Delta_0 - |\Delta|} z_2 + \sqrt{|\Delta|}z_3 + \theta_i^+ \right]
    \\
    &\times
    \Phi\left[ \sqrt{\Delta_0 - |\Delta|} z_1 + \sqrt{|\Delta|}z_3 + \mathrm{sgn}(\Delta) (\theta_i^+ +c_i)\right]
    +
    w_-
    \Phi\left[ \sqrt{\Delta_0 - |\Delta|} z_2 + \sqrt{|\Delta|}z_3 + (\theta_i^- +c_i) \right]
    \\
    &\times
    \Phi\left[ \sqrt{\Delta_0 - |\Delta|} z_1 + \sqrt{|\Delta|}z_3 + \mathrm{sgn}(\Delta) (\theta_i^-+c_i) \right]
    \Bigg\}\Bigg),
\end{align*}
where $w^\pm$ denote the relative contributions of a state being in either one of the two possible final $\theta$ states. These frequencies satisfy the normalization condition $w^+ + w^- = 1$.

We are particularly interested in the case $\alpha\beta>1$, which allows for further simplification. In this regime, we have $\theta^\pm = \pm \alpha$, leading to the final simplified form
\begin{align*}
V^\pm(\theta)
    &=
    -\frac{\Delta^2}{2}+ 
    \lim_{N\to\infty} \mathbf{E}_{N} \Bigg(
    \int_{-\infty}^{\infty}\int_{-\infty}^{\infty}\int_{-\infty}^{\infty}
    \mathrm{D}z_3 \mathrm{D}z_2 \mathrm{D}z_1
    \Bigg\{
    w_+
    \Phi\left[ \sqrt{\Delta_0 - |\Delta|} z_2 + \sqrt{|\Delta|}z_3 + (c_i +\alpha)\right]
    \\
    &\times
    \Phi\left[ \sqrt{\Delta_0 - |\Delta|} z_1 + \sqrt{|\Delta|}z_3 + (c_i +\mathrm{sgn}(\Delta) \alpha )\right]
    +
    w_-
    \Phi\left[ \sqrt{\Delta_0 - |\Delta|} z_1 + \sqrt{|\Delta|}z_3 +(c_i- \alpha) \right]
    \\
    &\times
    \Phi\left[ \sqrt{\Delta_0 - |\Delta|} z_2 + \sqrt{|\Delta|}z_3 +(c_i- \mathrm{sgn}(\Delta) \alpha )\right]
    \Bigg\}\Bigg) \\
    &=
    -\frac{\Delta^2}{2}+ 
    \lim_{N\to\infty} \mathbf{E}_{N} \Bigg(
    \int_{-\infty}^{\infty}\int_{-\infty}^{\infty}\int_{-\infty}^{\infty}
    \mathrm{D}z_3 \mathrm{D}z_2 \mathrm{D}z_1
    \Phi\left[ \sqrt{\Delta_0 - |\Delta|} z_2 + \sqrt{|\Delta|}z_3 + (c_i + \alpha) \right]
    \\
    &\times
    \Phi\left[ \sqrt{\Delta_0 - |\Delta|} z_1 + \sqrt{|\Delta|}z_3 + \mathrm{sgn}(\Delta) +(c_i +\alpha )\right]
    \Bigg).
\end{align*}

With this derivation, we have obtained the approximate long-time potential as shown in equation \eqref{eq:pmpotential}.

\twocolumngrid
\bibliographystyle{unsrt}
\bibliography{ProjectDMFT.bib}

\end{document}